\documentclass[journal]{IEEEtran} %
\pdfoutput=1
\usepackage[bookmarks=false]{hyperref}
\usepackage[utf8]{inputenc}
\usepackage{graphicx,url}
\usepackage{subcaption}
\usepackage{booktabs}
\graphicspath{ {./images/} }
\usepackage{multirow}
\usepackage{hhline}
\usepackage{graphicx}
\usepackage{amssymb,amsmath,mathtools}
\usepackage{amsthm}
\usepackage{cuted}
\newtheorem{theorem}{Theorem}

\usepackage{bm}
\usepackage[noadjust]{cite}

\usepackage{color,soul}
\usepackage[dvipsnames]{xcolor}
\hypersetup{
  colorlinks,
  linkcolor={blue!80!black},
  citecolor={blue!80!black},
  urlcolor={blue!80!black}
}
\usepackage{bbm}
\usepackage{lipsum,lineno}
\usepackage{float}
\makeatletter
\def\thm@space@setup{\thm@preskip=1mm
\thm@postskip=1mm}
\newif\if@restonecol
\makeatother
\usepackage{amsmath}
\usepackage{mathrsfs}
\usepackage[shortlabels]{enumitem}
\usepackage[titlenumbered,ruled,linesnumbered]{algorithm2e}

\newcommand\numeq[1]%
{\stackrel{\scriptscriptstyle(\mkern-1.5mu#1\mkern-1.5mu)}{=}}
\newcommand\numeqq[1]%
{\stackrel{\scriptscriptstyle(\mkern-1.5mu#1\mkern-1.5mu)}{\triangleq}}
\newcommand\numleq[1]%
{\stackrel{\scriptscriptstyle(\mkern-1.5mu#1\mkern-1.5mu)}{\leq}}
\newcommand\numgeq[1]%
{\stackrel{\scriptscriptstyle(\mkern-1.5mu#1\mkern-1.5mu)}{\geq}}
\newcommand\numimp[1]%
{\stackrel{\scriptscriptstyle(\mkern-1.5mu#1\mkern-1.5mu)}{\implies}}
\usepackage{fancyhdr} 
\usepackage{tabularx}

\SetCommentSty{mycommfont}
\usepackage{dsfont}
\usepackage{comment}
\usepackage{tikz,pgfplots}
\usepackage{pgfplots}
\definecolor{gray2}{HTML}{ededed}
\definecolor{gray3}{HTML}{F5F5F5}
\definecolor{gray4}{HTML}{e7f2f8}
\definecolor{gray10}{HTML}{780448}
\usetikzlibrary{shapes.geometric,backgrounds,patterns, trees}
\usetikzlibrary{3d,decorations.text,shapes.arrows,positioning,fit,backgrounds}
\usetikzlibrary{positioning, decorations.pathmorphing, shapes}
\usetikzlibrary{decorations.pathreplacing}
\usetikzlibrary{shapes.geometric,backgrounds,patterns, trees}
\usetikzlibrary{arrows.meta,
  bending,
  intersections,
  quotes,
  shapes.geometric}
\usetikzlibrary{automata, positioning}
\usepgfplotslibrary{fillbetween}
\usetikzlibrary{shapes,arrows}
\usetikzlibrary{arrows.meta}
\usetikzlibrary{positioning}
\tikzset{set/.style={draw,circle,inner sep=0pt,align=center}}
\usetikzlibrary{automata, positioning}
\usetikzlibrary{shapes,shadows}
\tikzstyle{abstractbox} = [draw=black, fill=white, rectangle,
inner sep=10pt, style=rounded corners, drop shadow={fill=black,
  opacity=1}]
\tikzstyle{abstracttitle} =[fill=white]
\tikzstyle{block} = [rectangle, draw,
text width=10.5em, text centered, rounded corners, minimum height=4em]
\tikzstyle{line} = [draw, -latex]
\usetikzlibrary{calc,positioning,shapes.geometric}
\usetikzlibrary{arrows.meta,arrows}
\DeclareMathOperator*{\argmax}{arg\,max}
\DeclareMathOperator*{\argmin}{arg\,min}

\DeclareMathOperator*{\maximize}{maximize}
\DeclareMathOperator*{\minimize}{minimize}

\usetikzlibrary{matrix}
\tikzstyle{cblue}=[circle, draw, thin,fill=cyan!20, scale=0.8]
\tikzstyle{qgre}=[rectangle, draw, thin,fill=green!20, scale=0.8]
\tikzstyle{rpath}=[dashed]
\tikzstyle{legend_isps}=[rectangle, rounded corners, thin,
fill=gray!20, text=blue, draw]

\tikzstyle{legend_overlay}=[rectangle, rounded corners, thin,
top color= white,bottom color=green!25,
minimum width=2.5cm, minimum height=0.8cm,
pinegreen]
\tikzstyle{legend_phytop}=[rectangle, rounded corners, thin,
top color= white,bottom color=cyan!25,
minimum width=2.5cm, minimum height=0.8cm,
royalblue]
\tikzstyle{legend_general}=[rectangle, rounded ckith'sers, thin,
top color= white,bottom color=lavander!25,
minimum width=2.5cm, minimum height=0.8cm,
violet]
\usetikzlibrary{matrix}

\colorlet{myRed}{red!20}
\tikzset{
  rows/.style 2 args={/utils/temp/.style={row ##1/.append style={nodes={#2}}},
    /utils/temp/.list={#1}},
  columns/.style 2 args={/utils/temp/.style={column ##1/.append style={nodes={#2}}},
    /utils/temp/.list={#1}}}
\usetikzlibrary{backgrounds,calc,shadings,shapes.arrows,shapes.symbols,shadows}
\definecolor{switch}{HTML}{006996}

\makeatletter
\pgfkeys{/pgf/.cd,
  parallelepiped offset x/.initial=2mm,
  parallelepiped offset y/.initial=2mm
}
\pgfdeclareshape{parallelepiped}
{
  \inheritsavedanchors[from=rectangle] 
  \inheritanchorborder[from=rectangle]
  \inheritanchor[from=rectangle]{north}
  \inheritanchor[from=rectangle]{north west}
  \inheritanchor[from=rectangle]{north east}
  \inheritanchor[from=rectangle]{center}
  \inheritanchor[from=rectangle]{west}
  \inheritanchor[from=rectangle]{east}
  \inheritanchor[from=rectangle]{mid}
  \inheritanchor[from=rectangle]{mid west}
  \inheritanchor[from=rectangle]{mid east}
  \inheritanchor[from=rectangle]{base}
  \inheritanchor[from=rectangle]{base west}
  \inheritanchor[from=rectangle]{base east}
  \inheritanchor[from=rectangle]{south}
  \inheritanchor[from=rectangle]{south west}
  \inheritanchor[from=rectangle]{south east}
  \backgroundpath{
    \southwest \pgf@xa=\pgf@x \pgf@ya=\pgf@y
    \northeast \pgf@xb=\pgf@x \pgf@yb=\pgf@y
    \pgfmathsetlength\pgfutil@tempdima{\pgfkeysvalueof{/pgf/parallelepiped
        offset x}}
    \pgfmathsetlength\pgfutil@tempdimb{\pgfkeysvalueof{/pgf/parallelepiped
        offset y}}
    \def\ppd@offset{\pgfpoint{\pgfutil@tempdima}{\pgfutil@tempdimb}}
    \pgfpathmoveto{\pgfqpoint{\pgf@xa}{\pgf@ya}}
    \pgfpathlineto{\pgfqpoint{\pgf@xb}{\pgf@ya}}
    \pgfpathlineto{\pgfqpoint{\pgf@xb}{\pgf@yb}}
    \pgfpathlineto{\pgfqpoint{\pgf@xa}{\pgf@yb}}
    \pgfpathclose
    \pgfpathmoveto{\pgfqpoint{\pgf@xb}{\pgf@ya}}
    \pgfpathlineto{\pgfpointadd{\pgfpoint{\pgf@xb}{\pgf@ya}}{\ppd@offset}}
    \pgfpathlineto{\pgfpointadd{\pgfpoint{\pgf@xb}{\pgf@yb}}{\ppd@offset}}
    \pgfpathlineto{\pgfpointadd{\pgfpoint{\pgf@xa}{\pgf@yb}}{\ppd@offset}}
    \pgfpathlineto{\pgfqpoint{\pgf@xa}{\pgf@yb}}
    \pgfpathmoveto{\pgfqpoint{\pgf@xb}{\pgf@yb}}
    \pgfpathlineto{\pgfpointadd{\pgfpoint{\pgf@xb}{\pgf@yb}}{\ppd@offset}}
  }
}

\makeatletter
\tikzset{anchor/.append code=\let\tikz@auto@anchor\relax,
  add font/.code=%
  \expandafter\def\expandafter\tikz@textfont\expandafter{\tikz@textfont#1},
  left delimiter/.style 2 args={append after command={\tikz@delimiter{south east}
      {south west}{every delimiter,every left delimiter,#2}{south}{north}{#1}{.}{\pgf@y}}}}
\tikzstyle{sms} = [rectangle callout, draw,very thick, rounded corners, minimum height=20pt]
\makeatletter
\tikzset{anchor/.append code=\let\tikz@auto@anchor\relax,
  add font/.code=%
  \expandafter\def\expandafter\tikz@textfont\expandafter{\tikz@textfont#1},
  left delimiter/.style 2 args={append after command={\tikz@delimiter{south east}
      {south west}{every delimiter,every left delimiter,#2}{south}{north}{#1}{.}{\pgf@y}}}}
\tikzstyle{sms} = [rectangle callout, draw,very thick, rounded corners, minimum height=20pt]
\usetikzlibrary{positioning,calc}

\tikzset{
  mybackground9/.style={execute at end picture={
      \begin{scope}[on background layer]
        \draw[black,fill=Black!5!Sepia!1,rounded corners=6ex] (current bounding box.south west)
        rectangle (current bounding box.north east);
        \node[draw,fill=white,ellipse,anchor=west,inner sep=1pt,minimum width=4ex] at (current bounding box.north
        west){#1};
      \end{scope}
    }},
}

\tikzset{
  mybackground10/.style={execute at end picture={
      \begin{scope}[on background layer]
        \draw[black] (current bounding box.south west)
        rectangle (current bounding box.north east);
        \node[draw,fill=white,ellipse,anchor=west,inner sep=1pt,minimum width=4ex] at (current bounding box.north
        west){#1};
      \end{scope}
    }},
}
\tikzset{
  mybackground21/.style={execute at end picture={
      \begin{scope}[on background layer]
        \draw[black, rounded corners=2ex, fill=gray2] (current bounding box.south west)
        rectangle (current bounding box.north east);
        \node[draw,fill=white,ellipse,anchor=west,inner sep=1pt,minimum width=4ex] at (current bounding box.north
        west){#1};
      \end{scope}
    }},
}

\tikzset{
  mybackground13/.style={execute at end picture={
      \begin{scope}[on background layer]
        \draw[black, fill=gray2, rounded corners=4ex] (current bounding box.south west)
        rectangle (current bounding box.north east);
        \node[draw,fill=white,ellipse,anchor=west,inner sep=1pt,minimum width=4ex] at (current bounding box.north
        west){#1};
      \end{scope}
    }},
}
\tikzset{
  mybackground14/.style={execute at end picture={
      \begin{scope}[on background layer]
        \draw[black, rounded corners=2ex] (current bounding box.south west)
        rectangle (current bounding box.north east);
        \node[draw,fill=white,ellipse,anchor=west,inner sep=1pt,minimum width=4ex] at (current bounding box.north
        west){#1};
      \end{scope}
    }},
}

\tikzset{
  mybackground11/.style={execute at end picture={
      \begin{scope}[on background layer]
        \draw[black, fill=Black!80!Sepia!9, rounded corners=6ex] (current bounding box.south west)
        rectangle (current bounding box.north east);
        \node[draw,fill=white,ellipse,anchor=west,inner sep=1pt,minimum width=4ex] at (current bounding box.north
        west){#1};
      \end{scope}
    }},
}

\tikzset{
  mybackground15/.style={execute at end picture={
      \begin{scope}[on background layer]
        \draw[black, fill=Black!80!Sepia!9, rounded corners=3ex] (current bounding box.south west)
        rectangle (current bounding box.north east);
        \node[draw,fill=white,ellipse,anchor=west,inner sep=1pt,minimum width=4ex] at (current bounding box.north
        west){#1};
      \end{scope}
    }},
}

\tikzset{
  mybackground12/.style={execute at end picture={
      \begin{scope}[on background layer]
        \draw[black, fill=Black!40!Emerald!30, rounded corners=3ex, line width=0.3mm] (current bounding box.south west)
        rectangle (current bounding box.north east);
      \end{scope}
    }},
}
\tikzset{
  mybackground18/.style={execute at end picture={
      \begin{scope}[on background layer]
        \draw[black, fill=gray3, rounded corners=3.5ex] (current bounding box.south west)
        rectangle (current bounding box.north east);
        \node[draw,fill=white,ellipse,anchor=west,inner sep=1pt,minimum width=4ex] at (current bounding box.north
        west){#1};
      \end{scope}
    }}
}

\tikzset{
  mybackground19/.style={execute at end picture={
      \begin{scope}[on background layer]
        \draw[black, fill=Black!10!gray4!60, rounded corners=1ex] (current bounding box.south west)
        rectangle (current bounding box.north east);
        west){#1};
      \end{scope}
    }},
}
\tikzset{
  mybackground27/.style={execute at end picture={
      \begin{scope}[on background layer]
        \draw[black, fill=gray2, rounded corners=2ex] (current bounding box.south west)
        rectangle (current bounding box.north east);
        \node[draw,fill=white,ellipse,anchor=west,inner sep=1pt,minimum width=4ex] at (current bounding box.north
        west){#1};
      \end{scope}
    }},
}
\tikzset{
  mybackground57/.style={execute at end picture={
      \begin{scope}[on background layer]
        \draw[black, fill=gray2, rounded corners=1ex] (current bounding box.south west)
        rectangle (current bounding box.north east);
      \end{scope}
    }},
}
\tikzset{
  mybackground48/.style={execute at end picture={
      \begin{scope}[on background layer]
        \draw[black, fill=gray3, rounded corners=4ex] (current bounding box.south west)
        rectangle (current bounding box.north east);
      \end{scope}
    }},
}

\tikzset{
  mybackground41/.style={execute at end picture={
      \begin{scope}[on background layer]
        \draw[black, rounded corners=2ex] (current bounding box.south west)
        rectangle (current bounding box.north east);
        \node[draw,fill=white,ellipse,anchor=west,inner sep=1pt,minimum width=4ex] at (current bounding box.north
        west){#1};
      \end{scope}
    }},
}

\tikzset{
  mybackground42/.style={execute at end picture={
      \begin{scope}[on background layer]
        \draw[black, rounded corners=2ex] (current bounding box.south west)
        rectangle (current bounding box.north east);
        \node[draw,fill=white,ellipse,anchor=east,inner sep=1pt,minimum width=4ex] at (current bounding box.north
        east){#1};
      \end{scope}
    }},
}
\tikzset{
  mybackground43/.style={execute at end picture={
      \begin{scope}[on background layer]
        \draw[Red, rounded corners=2ex, dashed] (current bounding box.south west)
        rectangle (current bounding box.north east);
        \node[draw,fill=white,ellipse,anchor=west,inner sep=1pt,minimum width=4ex] at (current bounding box.north
        west){#1};
      \end{scope}
    }},
}

\tikzset{
  mybackground44/.style={execute at end picture={
      \begin{scope}[on background layer]
        \draw[Red, rounded corners=2ex, dashed] (current bounding box.south west)
        rectangle (current bounding box.north east);
        \node[draw,fill=white,ellipse,anchor=east,inner sep=1pt,minimum width=4ex] at (current bounding box.north
        east){#1};
      \end{scope}
    }},
}

\tikzset{
  mybackground51/.style={execute at end picture={
      \begin{scope}[on background layer]
        \draw[black, rounded corners=2ex, fill=black!2] (current bounding box.south west)
        rectangle (current bounding box.north east);
        \node[draw,fill=white,ellipse,anchor=west,inner sep=1pt,minimum width=4ex] at (current bounding box.north
        west){#1};
      \end{scope}
    }},
}

\tikzset{
  mybackground52/.style={execute at end picture={
      \begin{scope}[on background layer]
        \draw[black, rounded corners=2ex, fill=black!2] (current bounding box.south west)
        rectangle (current bounding box.north east);
        \node[draw,fill=white,ellipse,anchor=east,inner sep=1pt,minimum width=4ex] at (current bounding box.north
        east){#1};
      \end{scope}
    }},
}

\tikzset{l3 switch/.style={
    parallelepiped,fill=switch, draw=white,
    minimum width=0.75cm,
    minimum height=0.75cm,
    parallelepiped offset x=1.75mm,
    parallelepiped offset y=1.25mm,
    path picture={
      \node[fill=white,
      circle,
      minimum size=6pt,
      inner sep=0pt,
      append after command={
        \pgfextra{
          \foreach \angle in {0,45,...,360}
          \draw[-latex,fill=white] (\tikzlastnode.\angle)--++(\angle:2.25mm);
        }
      }
      ]
      at ([xshift=-0.75mm,yshift=-0.5mm]path picture bounding box.center){};
    }
  },
  ports/.style={
    line width=0.3pt,
    top color=gray!20,
    bottom color=gray!80
  },
  rack switch/.style={
    parallelepiped,fill=white, draw,
    minimum width=1.25cm,
    minimum height=0.25cm,
    parallelepiped offset x=2mm,
    parallelepiped offset y=1.25mm,
    xscale=-1,
    path picture={
      \draw[top color=gray!5,bottom color=gray!40]
      (path picture bounding box.south west) rectangle
      (path picture bounding box.north east);
      \coordinate (A-west) at ([xshift=-0.2cm]path picture bounding box.west);
      \coordinate (A-center) at ($(path picture bounding box.center)!0!(path
      picture bounding box.south)$);
      \foreach \x in {0.275,0.525,0.775}{
        \draw[ports]([yshift=-0.05cm]$(A-west)!\x!(A-center)$)
        rectangle +(0.1,0.05);
        \draw[ports]([yshift=-0.125cm]$(A-west)!\x!(A-center)$)
        rectangle +(0.1,0.05);
      }
      \coordinate (A-east) at (path picture bounding box.east);
      \foreach \x in {0.085,0.21,0.335,0.455,0.635,0.755,0.875,1}{
        \draw[ports]([yshift=-0.1125cm]$(A-east)!\x!(A-center)$)
        rectangle +(0.05,0.1);
      }
    }
  },
  server/.style={
    parallelepiped,
    fill=white, draw,
    minimum width=0.35cm,
    minimum height=0.75cm,
    parallelepiped offset x=3mm,
    parallelepiped offset y=2mm,
    xscale=-1,
    path picture={
      \draw[top color=gray!5,bottom color=gray!40]
      (path picture bounding box.south west) rectangle
      (path picture bounding box.north east);
      \coordinate (A-center) at ($(path picture bounding box.center)!0!(path
      picture bounding box.south)$);
      \coordinate (A-west) at ([xshift=-0.575cm]path picture bounding box.west);
      \draw[ports]([yshift=0.1cm]$(A-west)!0!(A-center)$)
      rectangle +(0.2,0.065);
      \draw[ports]([yshift=0.01cm]$(A-west)!0.085!(A-center)$)
      rectangle +(0.15,0.05);
      \fill[black]([yshift=-0.35cm]$(A-west)!-0.1!(A-center)$)
      rectangle +(0.235,0.0175);
      \fill[black]([yshift=-0.385cm]$(A-west)!-0.1!(A-center)$)
      rectangle +(0.235,0.0175);
      \fill[black]([yshift=-0.42cm]$(A-west)!-0.1!(A-center)$)
      rectangle +(0.235,0.0175);
    }
  },
}
\pgfplotsset{compat=1.16}
\usetikzlibrary{calc, shadings, shadows, shapes.arrows}

\tikzset{%
  interface/.style={draw, rectangle, rounded corners, font=\LARGE\sffamily},
  ethernet/.style={interface, fill=yellow!50},
  serial/.style={interface, fill=green!70},
  speed/.style={sloped, anchor=south, font=\large\sffamily},
  route/.style={draw, shape=single arrow, single arrow head extend=4mm,
    minimum height=1.7cm, minimum width=3mm, white, fill=switch!20,
    drop shadow={opacity=.8, fill=switch}, font=\tiny}
}
\newcommand*{\shift}{1.3cm}
\newcommand*{\router}[1]{
  \begin{tikzpicture}
    \coordinate (ll) at (-3,0.5);
    \coordinate (lr) at (3,0.5);
    \coordinate (ul) at (-3,2);
    \coordinate (ur) at (3,2);
    \shade [shading angle=90, left color=switch, right color=white] (ll)
    arc (-180:-60:3cm and .75cm) -- +(0,1.5) arc (-60:-180:3cm and .75cm)
    -- cycle;
    \shade [shading angle=270, right color=switch, left color=white!50] (lr)
    arc (0:-60:3cm and .75cm) -- +(0,1.5) arc (-60:0:3cm and .75cm) -- cycle;
    \draw [thick] (ll) arc (-180:0:3cm and .75cm)
    -- (ur) arc (0:-180:3cm and .75cm) -- cycle;
    \draw [thick, shade, upper left=switch, lower left=switch,
    upper right=switch, lower right=white] (ul)
    arc (-180:180:3cm and .75cm);
    \node at (0,0.5){\color{blue!60!black}\Huge #1};
    \begin{scope}[yshift=2cm, yscale=0.28, transform shape]
      \node[route, rotate=45, xshift=\shift] {\strut};
      \node[route, rotate=-45, xshift=-\shift] {\strut};
      \node[route, rotate=-135, xshift=\shift] {\strut};
      \node[route, rotate=135, xshift=-\shift] {\strut};
    \end{scope}
  \end{tikzpicture}}

\makeatletter
\pgfdeclareradialshading[tikz@ball]{cloud}{\pgfpoint{-0.275cm}{0.4cm}}{%
  color(0cm)=(tikz@ball!75!white);
  color(0.1cm)=(tikz@ball!85!white);
  color(0.2cm)=(tikz@ball!95!white);
  color(0.7cm)=(tikz@ball!89!black);
  color(1cm)=(tikz@ball!75!black)
}

\tikzoption{cloud color}{\pgfutil@colorlet{tikz@ball}{#1}}
\makeatother

\tikzset{my cloud/.style={
    cloud, draw, aspect=2,
    fill={gray!15!white}
  }
}

\renewcommand\thetable{\arabic{table}}
\newcommand{\acro}[1]{\textsc{#1}\xspace}
\newcommand{\acros}[1]{\textsc{#1}s\xspace}
\newcommand{\acrop}[1]{\textsc{#1}\xspace}
\newcommand{\tpp}{\acrop{tp-2}}

\newcommand{\poposg}{\acro{po-posg}}

\newcommand{\mdp}{\acro{mdp}}
\newcommand{\pomdp}{\acro{pomdp}}

\newcommand{\dfsp}{\acro{dfsp}}
\newcommand{\nfsp}{\acro{nfsp}}

\newcommand{\pomdps}{\acros{pomdp}}
\newcommand{\idps}{\acro{idps}}

\newcommand{\mlr}{\acro{mlr}}

\newcommand{\sdn}{\acro{sdn}}

\newcommand{\cpu}{\acro{cpu}}
\newcommand{\cgroups}{\acros{cgroup}}

\newcommand{\openflow}{\acro{openflow}}

\newcommand{\ssh}{\acro{ssh}}

\newcommand{\irc}{\acro{irc}}
\newcommand{\smtp}{\acro{smtp}}
\newcommand{\snmp}{\acro{snmp}}
\newcommand{\ntp}{\acro{ntp}}
\newcommand{\spark}{\acro{spark}}

\newcommand{\ryu}{\acro{ryu}}
\newcommand{\debian}{\acro{debian}}
\newcommand{\jessie}{\acro{jessie}}
\newcommand{\wheezy}{\acro{wheezy}}
\newcommand{\samba}{\acro{samba}}
\newcommand{\tomcat}{\acro{tomcat}}

\newcommand{\dns}{\acro{dns}}
\newcommand{\snort}{\acro{snort}}
\newcommand{\http}{\acro{http}}

\newcommand{\vxlan}{\acro{vxlan}}

\newcommand{\netem}{\acro{netem}}
\newcommand{\apache}{\acro{apache}}
\newcommand{\ts}{\acro{ts}}
\newcommand{\mysql}{\acro{mysql}}
\newcommand{\tcpp}{\acro{tcp}}
\newcommand{\xmas}{\acro{xmas}}

\newcommand{\udp}{\acro{udp}}
\newcommand{\syn}{\acro{syn}}
\newcommand{\mongo}{\acro{mongodb}}
\newcommand{\postgres}{\acro{postgres}}
\newcommand{\telnet}{\acro{telnet}}
\newcommand{\cassandra}{\acro{cassandra}}
\newcommand{\ubuntu}{\acro{ubuntu}}
\newcommand{\vulscan}{\acro{vulscan}}
\newcommand{\hdfs}{\acro{hdfs}}
\newcommand{\ftp}{\acro{ftp}}

\newcommand{\ovs}{\acro{ovs}}
\newcommand{\dmz}{\acro{dmz}}
\newcommand{\rnd}{\acro{r\&d}}
\newcommand{\admin}{\acro{admin}}
\newcommand{\zone}{\acro{zone}}
\newcommand{\quarantine}{\acro{quarantine}}
\newcommand{\cve}{\acro{cve}}
\newcommand{\cwe}{\acro{cwe}}
\newcommand{\icmp}{\acro{icmp}}

\newcommand{\ppo}{\acro{ppo}}

\usepackage{etoolbox}           %
\makeatletter
\patchcmd{\@makecaption}
{\scshape}
{}
{}
{}
\makeatother

\allowdisplaybreaks
\begin{document}
\newcommand{\ssep}{\mid}
\bstctlcite{MyBSTcontrol}
\title{Scalable Learning of Intrusion Response\\
  through Recursive Decomposition}

\author{\IEEEauthorblockN{Kim Hammar \IEEEauthorrefmark{2}\IEEEauthorrefmark{3} and Rolf Stadler\IEEEauthorrefmark{2}\IEEEauthorrefmark{3}}

  \IEEEauthorblockA{\IEEEauthorrefmark{2}
    Division of Network and Systems Engineering, KTH Royal Institute of Technology, Sweden
  }\\
  \IEEEauthorblockA{\IEEEauthorrefmark{3} KTH Center for Cyber Defense and Information Security, Sweden \\
    Email: \{kimham, stadler\}@kth.se%
    \\
    \today
  }
}

\maketitle
\begin{abstract}
We study automated intrusion response for an IT infrastructure and formulate the interaction between an attacker and a defender as a partially observed stochastic game. To solve the game we follow an approach where attack and defense strategies co-evolve through reinforcement learning and self-play toward an equilibrium. Solutions proposed in previous work prove the feasibility of this approach for small infrastructures but do not scale to realistic scenarios due to the exponential growth in computational complexity with the infrastructure size. We address this problem by introducing a method that recursively decomposes the game into subgames with low computational complexity which can be solved in parallel. Applying optimal stopping theory we show that the best response strategies in these subgames exhibit threshold structures, which allows us to compute them efficiently. To solve the decomposed game we introduce an algorithm called Decompositional Fictitious Self-Play (\dfsp), which learns Nash equilibria through stochastic approximation. We evaluate the learned strategies in an emulation environment where real intrusions and response actions can be executed. The results show that the learned strategies approximate an equilibrium and that \dfsp significantly outperforms a state-of-the-art algorithm for a realistic infrastructure configuration.
\end{abstract}
\IEEEpeerreviewmaketitle
\section{Introduction}
A promising direction of recent research is to automatically find security strategies for an IT infrastructure through reinforcement learning methods, whereby the problem is formulated as a Markov decision problem and strategies are learned through simulation (see survey \cite{control_rl_reviews}). While encouraging results have been obtained following this approach (see e.g., \cite{han_yi_sdn} and \cite{hammar_stadler_tnsm}), key challenges remain. Most of the prior work, for example, follows a decision-theoretic formulation and aims at learning effective defender strategies against a static attacker with a fixed strategy \cite{han_yi_sdn,hammar_stadler_cnsm_20,hammar_stadler_cnsm_21,hammar_stadler_tnsm,csle_docs,noms_demo_preprint,Miehling_control_theoretic_approaches_summary,7011201,Kreidl2004FeedbackCA,miehling_attack_graph,miehling_control_security_3,deep_air,Iannucci2021AnIR,jakob_nyberg_1,rigaki2023cage,janisch2023nasimemu,foley2023inroads,hammar_stadler_cnsm_22,digital_twins_kim,hammar2022learning,al_shaer_book,hemberg_oreily_evo,causal_neil_agent,tabular_Q_andy,beyond_cage}. Only recently has the problem of learning effective security strategies against dynamic attackers been studied. This approach includes a game-theoretic framing, and the problem becomes one of learning Nash equilibria \cite{nework_security_alpcan,altman_jamming,hayel_ids,kamhoua2021game,hammar_stadler_game_23,tambe,5270307,radha_poove_game,umsonst2022bayesian,baras_gt,9833086}.
\begin{figure}
  \centering
  \scalebox{0.92}{
    \input{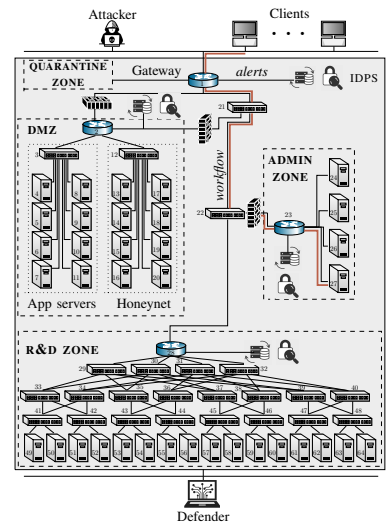}
  }
  \caption{The target infrastructure and the actors involved in the intrusion response use case.}
  \label{fig:use_case}
\end{figure}

Chief among the remaining challenges is the complexity of the formal model, resulting from the need to describe the target infrastructure with sufficient detail and at a realistic scale. Learning effective strategies with currently known methods is infeasible for most realistic use cases.

In this paper, we address the complexity challenge and present a scalable approach to automatically learn near-optimal defender strategies against dynamic attackers. We apply our approach to an \textit{intrusion response} use case that involves the IT infrastructure of an organization (see Fig. \ref{fig:use_case}). We formalize the use case as a partially observed stochastic game between two players -- the operator of the infrastructure, which we call the defender, and an attacker, which seeks to intrude on the infrastructure. To manage the complexity when formalizing the use case, we recursively decompose the game into simpler subgames, which allows detailed modeling of the infrastructure while keeping computational complexity low.

The decomposition involves three steps. First, we partition the infrastructure according to workflows that are isolated from each other. This allows us to decompose the game into \textit{independent subgames} (one per workflow) that can be solved in parallel. Second, the graph structure of a workflow allows us to decompose the workflow games into node subgames. We prove that these subgames have \textit{optimal substructure} \cite[Ch. 15]{cormen01introduction}, which means that a best response of the original game can be obtained from best responses of the node subgames. Third, we show that the problem of selecting \textit{which} response action to apply on a node can be separated from that of deciding \textit{when} to apply the action, which enables efficient learning of best responses through the application of \textit{optimal stopping} theory \cite{krishnamurthy_2016}. We use this insight to design an efficient reinforcement learning algorithm, called Decompositional Fictitious Self-Play (\dfsp), which allows scalable approximation of Nash equilibrium strategies.

\begin{figure}
  \centering
  \scalebox{0.77}{
    \input{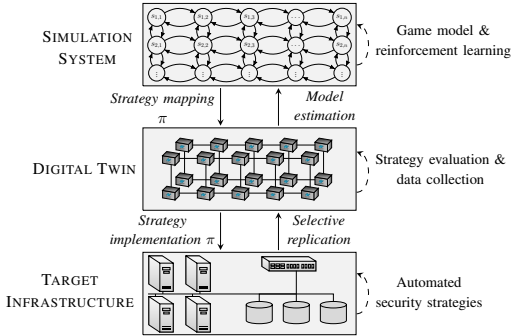}
  }
  \caption{Our framework for finding and evaluating intrusion response strategies \cite{hammar_stadler_tnsm,hammar_stadler_game_23,csle_docs}.}
  \label{fig:overview}
\end{figure}

Our method for learning the equilibrium strategies and evaluating them is based on a \textit{digital twin} of the target infrastructure, which we use to run attack scenarios and defender responses (see Fig. \ref{fig:overview}) \cite{hammar_stadler_tnsm,hammar_stadler_game_23,csle_docs}. Such runs produce system measurements and logs, from which we estimate infrastructure statistics. We then use these statistics to instantiate simulations of the infrastructure's dynamics and learn strategies through \dfsp. (Documentation of the software framework that implements the digital twin and the simulations is available at \cite{csle_docs,digital_twins_kim}; the source code is available at \cite{csle_source_code}; and a video demonstration is available at \cite{video_demonstration3}.)

We summarize the contributions in this paper as follows.
\begin{enumerate}
\item We formulate the intrusion response problem as a partially observed stochastic game and prove that, under assumptions often met in practice, the game decomposes into subgames whose best responses can be computed efficiently and in parallel.
\item We design \dfsp, an efficient reinforcement learning algorithm for approximating Nash equilibria of the decomposed game.
\item For a realistic use case, we evaluate the learned response strategies against network intrusions on a digital twin.
\end{enumerate}
\section{Related Work}
Networked systems found in engineering and science often exhibit a modular topological structure that can be exploited for designing control algorithms \cite{decomp_game_2}. System decomposition for the purpose of automatic control was first suggested by Šiljak in 1978 \cite{siljak1978large} and approaches based on decomposition, such as divide and conquer, layering, and hierarchical structuring are well established in the design of large-scale systems, a notable example being the Internet \cite{the_osi_reference_model}. Similar decomposition methods are frequently used in robotics and multi-agent systems, as exemplified by the subsumption architecture \cite{subsumption_arch_brooks}. Within the fields of decision- and game-theory, decomposition is studied in the context of factored decision processes \cite{decomposed_mdps,factored_mdps_2,NIPS1997_90db9da4,transition_indep_mdps}, distributed decision processes \cite{nd_pomdp_1}, factored games \cite{quanyan_decompose,oliehoek2014exploiting}, and graph-structured games \cite{kearns_graphical_games}.

Decomposition as a means to automate intrusion response has been studied first in \cite{quanyan_decompose,miehling_control_security_2,6759943,5608747}. The work in \cite{quanyan_decompose} formulates the interaction between a defender and an attacker on a cyber-physical infrastructure as a factored Markov game and introduces a decomposition based on linear programming. Following a similar approach, the work in \cite{6759943} studies a Markov game formulation and shows that a multi-stage game can be decomposed into a sequence of one-stage games. In a separate line of work, \cite{miehling_control_security_2} models intrusion response as a minimax control problem and develops a heuristic decomposition based on clustering and influence graphs. This approach resembles the work in \cite{5608747}, which studies a factored decision process and proposes a hierarchical decomposition.

In all of the above works, decomposition is key to obtain effective strategies for large-scale systems. Compared to our work, some of them propose decomposition methods without optimal substructure \cite{miehling_control_security_2}, others do not consider partial observability \cite{quanyan_decompose,6759943}, or dynamic attackers \cite{5608747}. Most importantly, all of the above works evaluate the obtained strategies in a simulation environment. They do not perform evaluation in an emulation environment as we report in this paper, which gives higher confidence that the strategies are effective on the target infrastructure.

For a comprehensive review of prior research on automated intrusion response (beyond work that use decomposition), see \cite[\S VII]{hammar_stadler_game_23}.
\section{The Intrusion Response Use Case}\label{sec:use_case}
We consider an intrusion response use case that involves the IT infrastructure of an organization. The operator of this infrastructure, which we call the defender, takes measures to protect it against an attacker while providing services to a client population (see Fig. \ref{fig:use_case}). The infrastructure is segmented into \textit{zones} with virtual \textit{nodes} that run network services. Services are realized by \textit{workflows} that are accessed by clients through a gateway, which also is open to the attacker.

\begin{figure}
  \centering
  \scalebox{1}{
    \input{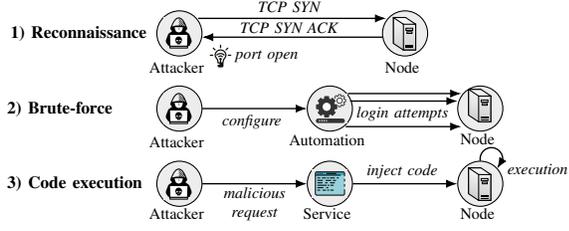}
  }
  \caption{Attacker actions: (\textit{i}) reconnaissance actions; (\textit{ii}) brute-force attacks; and (\textit{iii}) code execution attacks.}
  \label{fig:attacker_actions}
\end{figure}

The attacker's goal is to intrude on the infrastructure, compromise nodes, and disrupt workflows. It can take three types of actions to achieve this goal: (\textit{i}) reconnaissance; (\textit{ii}) brute-force attacks; and (\textit{iii}) exploits (see Fig. \ref{fig:attacker_actions}).

The defender continuously monitors the infrastructure through accessing and analyzing intrusion detection alerts and other statistics. It can take four types of defensive actions to respond to possible intrusions: (\textit{i}) migrate nodes between zones; (\textit{ii}) redirect or block network flows; (\textit{iii}) shut down nodes; and (\textit{iv}) revoke access to nodes (see Fig. \ref{fig:defender_actions}). When deciding between these actions, the defender must balance two conflicting objectives: maximize workflow utility towards its clients and minimize the cost of intrusion.
\section{Formalizing the Intrusion Response Problem}\label{sec:system_model}
We formalize the above use case as an optimization problem where the goal is to select an optimal sequence of defender actions in response to a sequence of attacker actions. We assume a dynamic attacker, which leads to a game-theoretic formulation of the intrusion response problem. The game is played on the IT infrastructure, which we model as a discrete-time dynamical system whose evolution depends on the actions by the attacker and the defender. Both actors have partial observability of the system state, and their observations depend on the traffic generated by clients requesting service, which we assume can be described by a stationary process.

\vspace{1mm}

\noindent\textbf{Notations.} Boldface lower case letters (e.g., $\mathbf{x}$) denote row vectors and upper case calligraphic letters (e.g., $\mathcal{V}$) represent sets. The set of probability distributions over $\mathcal{V}$ is written as $\Delta(\mathcal{V})$. A random variable is denoted by upper case (e.g., $X$) and a random vector is denoted by boldface (e.g., $\mathbf{X}=(X_1,\hdots,X_n)$). $\mathbb{P}$ is the probability measure and the expectation of $f$ with respect to $X$ is expressed as $\mathbb{E}_X[f]$. When $f$ includes many random variables that depend on $\pi$ we simplify the notation to $\mathbb{E}_{\pi}[f]$. We use $x \sim f$ to mean that $x$ is sampled from $f$ and sometimes write $\mathbb{P}[x|z,y]$ instead of $\mathbb{P}[X=x|Z=z,Y=y]$ when $X,Z,Y$ are clear from the context. Symbols used throughout the paper are listed in Table \ref{tab:notation}.
\subsection{Modeling the Infrastructure and Services}\label{sec:infra_moel}
Following the description in \S \ref{sec:use_case}, we consider an IT infrastructure with application servers connected by a communication network that is segmented into zones (see Fig. \ref{fig:use_case}). Overlaid on this physical infrastructure is a virtual infrastructure with tree-structure that includes \textit{nodes}, which collectively offer services to clients.

\begin{figure}
  \centering
  \scalebox{0.92}{
    \input{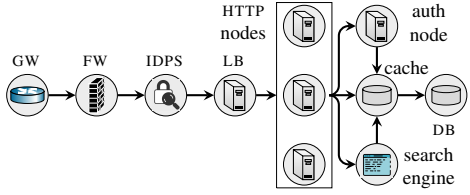}
  }
  \caption{Dependency graph of a workflow consisting of a tree of virtual network functions and microservices; \textsc{fw}, \textsc{lb}, and \textsc{idps} are acronyms for firewall, load balancer, and intrusion detection and prevention system, respectively.}
  \label{fig:workflow_dependency_diagram}
\end{figure}

A service is modeled as a \textit{workflow}, which comprises a set of interdependent nodes. A dependency between two nodes reflects information exchange through service invocations. We assume that each node belongs to exactly one workflow. As an example of a virtual infrastructure, we can think of a microservice architecture where a workflow is defined as a tree of microservices (see Fig. \ref{fig:workflow_dependency_diagram}).

\vspace{1mm}

\noindent\textbf{Infrastructure.} We model the virtual infrastructure as a (finite) directed graph $\mathcal{G} \triangleq \langle \{\mathrm{gw}\}\cup \mathcal{V}, \mathcal{E} \rangle$. The graph has a tree structure and is rooted at the gateway $\mathrm{gw}$. Each node $i \in \mathcal{V}$ has three state variables. $v_{i,t}^{(\mathrm{R})}$ represents the reconnaissance state and realizes the binary random variable $V_{i,t}^{(\mathrm{R})}$. $v_{i,t}^{(\mathrm{R})}=1$ if the attacker has discovered the node, $0$ otherwise. $v_{i,t}^{(\mathrm{I})}$ represents the intrusion state and realizes the binary random variable $V_{i,t}^{(\mathrm{I})}$. $v_{i,t}^{(\mathrm{I})}=1$ if the attacker has compromised the node, $0$ otherwise. Lastly, $v^{(\mathrm{Z})}_{i,t}$ indicates the zone in which the node resides and realizes the random variable $V^{(\mathrm{Z})}_{i,t}$. We call a node \textit{active} if it is functional as part of a workflow (denoted $\alpha_{i,t}=1$). Due to a defender action (e.g., a shut down) a node $i \in \mathcal{V}$ may become inactive (i.e., $\alpha_{i,t}=0$).

\vspace{1mm}

\noindent\textbf{Workflows.} We model a workflow $\mathbf{w} \in \mathcal{W}$ as a subtree $\mathcal{G}_{\mathbf{w}} \triangleq \langle \{\mathrm{gw}\}\cup \mathcal{V}_{\mathbf{w}}, \mathcal{E}_{\mathbf{w}} \rangle$ of the infrastructure graph. Workflows do not overlap except for the gateway which belongs to all workflows.

\begin{figure}
  \centering
  \scalebox{0.8}{
    \input{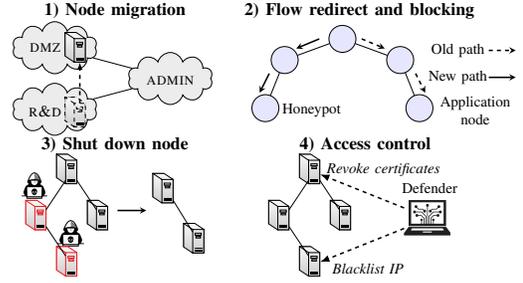}
  }
  \caption{Defender actions: (\textit{i}) migrate a node between two zones; (\textit{ii}) redirect or block traffic flows to a node; (\textit{iii}) shut down a node; and (\textit{iv}) revoke access to a node.}
  \label{fig:defender_actions}
\end{figure}
\subsection{Modeling Actors}
The intrusion response use case involves three types of actors: an attacker, a defender, and clients (see Fig. \ref{fig:use_case}).

\vspace{1mm}

\noindent\textbf{Attacker.} At each time $t$, the attacker takes an action $\mathbf{a}^{(\mathrm{A})}_t$, which is defined as the composition of the local actions on all nodes $\mathbf{a}^{(\mathrm{A})}_t \triangleq (\mathbf{a}^{(\mathrm{A})}_{1,t}, \hdots, \mathbf{a}^{(\mathrm{A})}_{|\mathcal{V}|,t}) \in \mathcal{A}_{\mathrm{A}}$, where $\mathcal{A}_{\mathrm{A}}$ is finite. A local action is either the null action (denoted with $\bot$) or an offensive action (see examples in Fig. \ref{fig:attacker_actions}). An offensive action on a node $i$ may change the reconnaissance state $v^{(\mathrm{R})}_{i,t}$ or the intrusion state $v^{(\mathrm{I})}_{i,t}$. A node $i$ can only be compromised if it is discovered, i.e., if $v^{(\mathrm{R})}_{i,t}=1$. We express this constraint as $\mathbf{a}^{(\mathrm{A})}_t \in \mathcal{A}_{\mathrm{A}}(\mathbf{s}^{(\mathrm{A})}_t)$.

The attacker state $\mathbf{S}^{(\mathrm{A})}_t \triangleq \big(V^{(\mathrm{I})}_{i,t},V^{(\mathrm{R})}_{i,t}\big)_{i \in \mathcal{V}} \in \mathcal{S}_{\mathrm{A}}$ evolves as
\begin{align}
\mathbf{s}^{(\mathrm{A})}_{t+1} \sim f_{\mathrm{A}}\big(\cdot \mid \mathbf{s}^{(\mathrm{A})}_t, \mathbf{a}^{(\mathrm{A})}_t,\mathbf{a}^{(\mathrm{D})}_t\big)\label{eq:transition_atc}
\end{align}
where $\mathbf{S}^{(\mathrm{A})}_t$, $\mathbf{A}^{(\mathrm{A})}_t$, and $\mathbf{A}^{(\mathrm{D})}_t$ are random vectors with realizations $\mathbf{s}^{(\mathrm{A})}_t$, $\mathbf{a}^{(\mathrm{A})}_t$, and $\mathbf{a}^{(\mathrm{D})}_t$. ($\mathbf{A}^{(\mathrm{D})}_t$ represents the defender action at time $t$.)

\vspace{1mm}

\noindent\textbf{Defender.} At each time $t$, the defender takes an action $\mathbf{a}^{(\mathrm{D})}_t$, which is defined as the composition of the local actions on all nodes $\mathbf{a}^{(\mathrm{D})}_t \triangleq (\mathbf{a}^{(\mathrm{D})}_{1,t}, \hdots, \mathbf{a}^{(\mathrm{D})}_{|\mathcal{V}|,t}) \in \mathcal{A}_{\mathrm{D}}$, where $\mathcal{A}_{\mathrm{D}}$ is finite. A local action is either a defensive action or the null action $\bot$ (see examples in Fig. \ref{fig:defender_actions}). Each defensive action $\mathbf{a}^{(\mathrm{D})}_{i,t}\neq \bot$ leads to $\mathbf{S}^{(\mathrm{A})}_{i,t+1}=(0,0)$ and may affect $V^{(\mathrm{Z})}_{i,t+1}$.

The defender state $\mathbf{S}^{(\mathrm{D})}_t \triangleq \big(V^{(\mathrm{Z})}_{i,t}\big)_{i \in \mathcal{V}} \in \mathcal{S}_{\mathrm{D}}$ evolves according to
\begin{align}
\mathbf{s}^{(\mathrm{D})}_{t+1} \sim f_{\mathrm{D}}\big(\cdot \mid \mathbf{s}^{(\mathrm{D})}_t, \mathbf{a}^{(\mathrm{D})}_t\big)\label{eq:transition_def}
\end{align}
where $\mathbf{s}^{(\mathrm{D})}_{t}$ and $\mathbf{a}^{(\mathrm{D})}_{t}$ realize the random vectors $\mathbf{S}^{(\mathrm{D})}_{t}$ and $\mathbf{A}^{(\mathrm{D})}_t$.

\vspace{1mm}

\noindent\textbf{Clients.} Clients consume services of the infrastructure by accessing workflows. We model client behavior through stationary stochastic processes, which affect the observations available to the attacker and the defender.
\subsection{Observability and Strategies}
At each time $t$, the defender and the attacker both observe $\mathbf{o}_t \triangleq \left(\mathbf{o}_{1,t},\hdots,\mathbf{o}_{|\mathcal{V}|,t}\right) \in \mathcal{O}$, where $\mathcal{O}$ is finite. (In our use case  $\mathbf{o}_t$ relates to the number of \idps alerts per node.) $\mathbf{o}_t$ is drawn from the random vector $\mathbf{O}_t \triangleq (\mathbf{O}_{1,t},\hdots,\mathbf{O}_{|\mathcal{V}|,t})$ whose marginal distributions $Z_{\mathbf{O}_1}, \hdots, Z_{\mathbf{O}_{|\mathcal{V}|}}$ are stationary and conditionally independent given $\mathbf{S}_{i,t}\triangleq(\mathbf{S}^{(\mathrm{D})}_{i,t},\mathbf{S}^{(\mathrm{A})}_{i,t})$. (Note that $Z_{\mathbf{O}_i}$ depends on the traffic generated by clients.) As a consequence, the joint conditional distribution $Z$ is given by
\begin{align}
Z\big(\mathbf{O}_{t}=\mathbf{o} \mid \mathbf{s}_{t}\big)=\prod^{|\mathcal{V}|}_{i=1} Z_{\mathbf{O}_i}\big(\mathbf{O}_{i,t}=\mathbf{o}_{i} \mid \mathbf{s}_{i,t}\big) && \forall \mathbf{o} \in \mathcal{O}\label{eq:obs_1}
\end{align}

The sequence of observations and states at times $1,\hdots,t$ forms the histories $\mathbf{h}^{(\mathrm{D})}_t\in \mathcal{H}_{\mathrm{D}}$ and $\mathbf{h}^{(\mathrm{A})}_t\in \mathcal{H}_{\mathrm{A}}$. These histories are realizations of the random vectors $\mathbf{H}^{(\mathrm{D})}_t\allowbreak\triangleq\allowbreak(\mathbf{S}^{(\mathrm{D})}_1,\allowbreak\mathbf{A}_1^{(\mathrm{D})},\allowbreak\mathbf{O}_1,\allowbreak\hdots,\allowbreak\mathbf{A}_{t-1}^{(\mathrm{D})},\allowbreak\mathbf{S}^{(\mathrm{D})}_t,\allowbreak\mathbf{O}_t)$ and $\mathbf{H}^{(\mathrm{A})}_t\allowbreak\triangleq \allowbreak(\mathbf{S}_1^{(\mathrm{A})},\allowbreak\mathbf{A}_1^{(\mathrm{A})},\allowbreak\mathbf{O}_1,\allowbreak\hdots,\allowbreak\mathbf{A}_{t-1}^{(\mathrm{A})},\allowbreak\mathbf{S}_t^{(\mathrm{A})},\allowbreak\mathbf{O}_t)$. Based on their respective histories, the defender and the attacker select actions, which define the defender strategy $\pi_{\mathrm{D}}\in \Pi_{\mathrm{D}}: \mathcal{H}_{\mathrm{D}} \rightarrow \Delta(\mathcal{A}_{\mathrm{D}})$ and the attacker strategy $\pi_{\mathrm{A}} \in \Pi_{\mathrm{A}}: \mathcal{H}_{\mathrm{A}} \rightarrow \Delta(\mathcal{A}_{\mathrm{A}})$.
\subsection{The Intrusion Response Problem}\label{sec:ir_problem}
When selecting the strategy $\pi_{\mathrm{D}}$ the defender must balance two conflicting objectives: maximize the workflow utility towards its clients and minimize the cost of intrusion. The weight $\eta \geq 0$ controls the trade-off between these two objectives, which results in the bi-objective
\begin{align}
  &J \triangleq \sum_{t=1}^{\infty}\gamma^{t-1} \Biggl(\sum_{\mathbf{w} \in \mathcal{W}}\sum_{i \in \mathcal{V}_{\mathbf{w}}}\underbrace{\eta u^{(\mathrm{W})}_{i,t}}_{\text{workflows utility}} - \underbrace{c_{i,t}^{(\mathrm{I})}}_{\text{intrusion cost}}\Biggr) \label{eq:objective_fun}
\end{align}
where $\gamma \in [0,1)$ is a discount factor, $c_{i,t}^{(\mathrm{I})}$ is the intrusion cost associated with node $i$ at time $t$, and $u^{(\mathrm{W})}_{i,t}$ expresses the workflow utility associated with node $i$ at time $t$. For this paper, we assume that $u^{(\mathrm{W})}_{i,t}$ is proportional to the number of active nodes in the subtree rooted at $i$ and that $c_{i,t}^{(\mathrm{I})}= v^{(\mathrm{I})}_{i,t} + c^{(\mathrm{A})}(\mathbf{a}^{(\mathrm{D})}_{i,t})$, where $c^{(\mathrm{A})}$ is a non-negative function.

Given (\ref{eq:objective_fun}) and an attacker strategy $\pi_{\mathrm{A}}$, the intrusion response problem can be stated as
\begin{subequations}\label{eq:formal_problem}
  \begin{align}
    \maximize_{\pi_{\mathrm{D}} \in \Pi_{\mathrm{D}}} \quad &\mathbb{E}_{(\pi_{\mathrm{D}}, \pi_{\mathrm{A}})}\left[J\right]\\
    \text{subject to} \quad &\mathbf{s}^{(\mathrm{D})}_{t+1} \sim f_{\mathrm{D}}\big(\cdot \mid \mathbf{s}^{(\mathrm{D})}_t, \mathbf{a}^{(\mathrm{D})}_t\big) && \forall t \label{eq:dynamics_constraint}\\
                                                     &\mathbf{s}^{(\mathrm{A})}_{t+1} \sim f_{\mathrm{A}}\big(\cdot \mid \mathbf{s}^{(\mathrm{A})}_t, \mathbf{a}^{(\mathrm{A})}_t, \mathbf{a}^{(\mathrm{D})}_t\big) && \forall t \label{eq:dynamics_constraint_attacker}\\
                                                     &\mathbf{o}_{t+1} \sim Z\big(\cdot \mid \mathbf{s}^{(\mathrm{D})}_{t+1}, \mathbf{s}^{(\mathrm{A})}_{t+1}) && \forall t \label{eq:obs_constraint}\\
                                                            &\mathbf{a}^{(\mathrm{A})}_{t} \sim \pi_{\mathrm{A}}\big(\cdot \mid \mathbf{h}^{(\mathrm{A})}_t\big) && \forall t\label{eq:attacker_strategy_constraint}\\
                                                            &\mathbf{a}^{(\mathrm{D})}_{t} \sim \pi_{\mathrm{D}}\big(\cdot \mid \mathbf{h}^{(\mathrm{D})}_t\big) && \forall t\label{eq:defender_strategy_constraint}\\
                                                            &\mathbf{s}^{(\mathrm{D})}_{t} \in \mathcal{S}_{\mathrm{D}}, \quad \mathbf{s}^{(\mathrm{A})}_{t} \in \mathcal{S}_{\mathrm{A}},\quad \mathbf{o} \in \mathcal{O} && \forall t\label{eq:domain_constraint_1} \\
    &\mathbf{a}^{(\mathrm{D})}_{t} \in \mathcal{A}_{\mathrm{D}}, \quad \mathbf{a}^{(\mathrm{A})}_{t} \in \mathcal{A}_{\mathrm{A}}(\mathbf{s}^{(\mathrm{A})}_t) && \forall t \label{eq:domain_constraint_2}\\
                                                            &\mathbf{s}^{(\mathrm{A})}_1 \sim \mathbf{b}^{(\mathrm{A})}_1 \label{eq:init_dist_a}\\
                                                            &\mathbf{s}^{(\mathrm{D})}_1 \sim \mathbf{b}^{(\mathrm{D})}_1\label{eq:init_dist_d}
  \end{align}
\end{subequations}
where $\mathbb{E}_{(\pi_{\mathrm{D}}, \pi_{\mathrm{A}})}$ denotes the expectation over the random vectors $(\mathbf{H}^{(\mathrm{D})}_t,\mathbf{H}^{(\mathrm{A})}_t)_{t\in \{1,2,\hdots\}}$ when following the strategy profile $(\pi_{\mathrm{D}},\pi_{\mathrm{A}})$; (\ref{eq:dynamics_constraint})--(\ref{eq:dynamics_constraint_attacker}) are the dynamics constraints; (\ref{eq:obs_constraint}) describes the observations; (\ref{eq:attacker_strategy_constraint})--(\ref{eq:defender_strategy_constraint}) capture the actions; (\ref{eq:domain_constraint_1})--(\ref{eq:domain_constraint_2}) are the domain constraints; and (\ref{eq:init_dist_a})--(\ref{eq:init_dist_d}) define the initial state distributions. (As a maximizer of (\ref{eq:formal_problem}) exists (see Thm. \ref{thm:equilibrium}), we write $\max$ instead of $\sup$ throughout this paper).

Solving (\ref{eq:formal_problem}) yields an optimal defender strategy against a \textit{static} attacker with a fixed strategy. Note that this defender strategy is generally not optimal against a different attacker strategy. For this reason, we aim to find a defender strategy that maximizes the minimum value of $J$ (\ref{eq:objective_fun}) across all possible attacker strategies. This objective can be formally expressed as a maxmin problem:
  \begin{align}
    \maximize_{\pi_{\mathrm{D}} \in \Pi_{\mathrm{D}}}\minimize_{\pi_{\mathrm{A}} \in \Pi_{\mathrm{A}}}&\text{ } \text{ }\mathbb{E}_{(\pi_{\mathrm{D}}, \pi_{\mathrm{A}})}\left[J\right] \text{ subject to (\ref{eq:dynamics_constraint})--(\ref{eq:init_dist_d})} \label{eq:game_def_maxmin}
  \end{align}
Solving (\ref{eq:game_def_maxmin}) corresponds to finding a Nash equilibrium \cite[Eq. 1]{nash51} of a two-player game. Hence the problem of solving (\ref{eq:game_def_maxmin}) can be analyzed through game theory.
\section{The Intrusion Response Game}\label{sec:game_formulation}
The maxmin problem in (\ref{eq:game_def_maxmin}) defines a stationary, finite, and zero-sum Partially Observed Stochastic Game with Public Observations (a \poposg) \cite[Def. 1]{po_posgs_horak_bosansky}\cite{Shapley1095}:
\begin{align}
\Gamma = \langle \mathcal{N}, (\mathcal{S}_{\mathrm{k}},\mathcal{A}_{\mathrm{k}},f_{\mathrm{k}},\mathbf{b}^{(\mathrm{k})}_1)_{\substack{\mathrm{k} \in \mathcal{N}}}, u, \gamma, \mathcal{O}, Z \rangle \label{def:game}
\end{align}
The game $\Gamma$ has two players $\mathcal{N}=\{\mathrm{D},\mathrm{A}\}$ with $\mathrm{D}$ being the defender and $\mathrm{A}$ being the attacker. $(\mathcal{S}_{\mathrm{k}})_{\mathrm{k} \in \mathcal{N}}$ are the state spaces, $(\mathcal{A}_{\mathrm{k}})_{\mathrm{k} \in \mathcal{N}}$ are the action spaces, and $\mathcal{O}$ is observation space (as defined in \S \ref{sec:system_model}). The transition functions $(f_{\mathrm{k}})_{\mathrm{k} \in \mathcal{N}}$ are defined by (\ref{eq:dynamics_constraint})--(\ref{eq:dynamics_constraint_attacker}), the observation function $Z$ is defined in (\ref{eq:obs_1}), and the utility function $u(\mathbf{s}_t,\mathbf{a}^{(\mathrm{D})}_{t})$ is the expression within brackets in (\ref{eq:objective_fun}). $(\mathbf{b}^{(\mathrm{k})}_1)_{\mathrm{k} \in \mathcal{N}}$ are the state distributions at $t=1$ and $\gamma$ is the discount factor in (\ref{eq:objective_fun}).

\vspace{1mm}

\noindent\textbf{Game play.} When the game starts at $t=1$, $\mathbf{s}^{(\mathrm{D})}_1$ and $\mathbf{s}_1^{(\mathrm{A})}$ are sampled from $\mathbf{b}^{(\mathrm{D})}_1$ and $\mathbf{b}^{(\mathrm{A})}_1$. A play of the game proceeds in time-steps $t=1,2,\hdots$. At each time $t$, the defender observes $\mathbf{h}^{(\mathrm{D})}_t$ and the attacker observes $\mathbf{h}^{(\mathrm{A})}_t$. Based on these histories, both players select actions according to their respective strategies, i.e., $\mathbf{a}^{(\mathrm{D})}_{t} \sim \pi_{\mathrm{D}}(\cdot \mid \mathbf{h}^{(\mathrm{D})}_t)$ and $\mathbf{a}^{(\mathrm{A})}_{t} \sim \pi_{\mathrm{A}}(\cdot \mid \mathbf{h}^{(\mathrm{A})}_t)$. As a result of these actions, five events occur at time $t+1$: (\textit{i}) $\mathbf{o}_{t+1}$ is sampled from $Z$; (\textit{ii}) $\mathbf{s}_{t+1}^{(\mathrm{D})}$ is sampled from $f_{\mathrm{D}}$; (\textit{iii}) $\mathbf{s}_{t+1}^{(\mathrm{A})}$ is sampled from $f_{\mathrm{A}}$; (\textit{iv}) the defender receives the utility $u(\mathbf{s}_t,\mathbf{a}^{(\mathrm{D})}_{t})$; and (\textit{v}) the attacker receives the utility $-u(\mathbf{s}_t,\mathbf{a}^{(\mathrm{D})}_{t})$.

\vspace{1mm}

\noindent\textbf{Belief states.} Based on their histories $\mathbf{h}^{(\mathrm{D})}_t$ and $\mathbf{h}^{(\mathrm{A})}_t$, both players form beliefs about the unobservable components of the state $\mathbf{s}_t$, which are expressed through the belief states $\mathbf{b}^{(\mathrm{D})}_t(\mathbf{s}^{(\mathrm{A})}_t)\triangleq\mathbb{P}[\mathbf{s}^{(\mathrm{A})}_t \mid \mathbf{H}^{(\mathrm{D})}_t=\mathbf{h}^{(\mathrm{D})}_t]$ and $\mathbf{b}^{(\mathrm{A})}_t(\mathbf{s}^{(\mathrm{D})}_t)\triangleq\mathbb{P}[\mathbf{s}^{(\mathrm{D})}_t \mid \mathbf{H}^{(\mathrm{A})}_t=\mathbf{h}^{(\mathrm{A})}_t]$.

The belief states are are realizations of $\mathbf{B}^{(\mathrm{D})}_t$ and $\mathbf{B}^{(\mathrm{A})}_t$ and are updated at each time $t>1$ via \cite[Eq. 1]{po_posgs_horak_bosansky}
\begin{align}
  &\mathbf{b}^{(\mathrm{k})}_{t}(\mathbf{s}^{(\mathrm{-k})}_{t}) = C_{\mathrm{k}} \sum_{\mathbf{s}^{(\mathrm{-k})}_{t-1} \in \mathcal{S}_{\mathrm{-k}}}\sum_{\mathbf{a}^{(\mathrm{-k})}_{t-1} \in \mathcal{A}_{\mathrm{-k}}(\mathbf{s}_t)}\mathbf{b}^{(\mathrm{k})}_{t-1}(\mathbf{s}^{(\mathrm{-k})}_{t-1})\cdot \label{eq:irs_belief_upd}\\
  &\pi^{(s)}_{\mathrm{-k}}(\mathbf{a}^{(\mathrm{-k})}_{t-1}\mid \mathbf{s}^{(\mathrm{-k})}_{t-1})Z(\mathbf{o}_{t} \mid \mathbf{s}^{(\mathrm{D})}_{t}, \mathbf{a}^{(\mathrm{A})}_{t-1}) f_{\mathrm{-k}}(\mathbf{s}^{(\mathrm{-k})}_{t} \mid \mathbf{s}^{(\mathrm{-k})}_{t-1}, \mathbf{a}_{t-1})\nonumber
\end{align}
where $\mathrm{k} \in \{\mathrm{D}, \mathrm{A}\}$, $C_{\mathrm{k}}=1/\mathbb{P}[\mathbf{o}_{t}\mid \mathbf{s}^{(\mathrm{k})}_{t}, \mathbf{a}^{(\mathrm{k})}_{t-1},\pi_{\mathrm{-k}}, \mathbf{b}^{(\mathrm{k})}_{t-1}]$ is a normalizing factor that makes the components of $\mathbf{b}^{(\mathrm{k})}_{t}$ sum to $1$ and $\pi^{(s)}_{\mathrm{-k}}: \mathcal{S}_{\mathrm{-k}} \rightarrow \Delta(\mathcal{A}_{\mathrm{-k}})$ is the stage strategy for the opponent, i.e., the strategy of the opponent in the current stage only, which is assumed known to both players at each stage \cite{po_posgs_horak_bosansky}.

The initial beliefs at $t=1$ are the degenerate distributions $\mathbf{b}^{(\mathrm{D})}_1(\mathbf{0}_{2|\mathcal{V}|})=1$ and $\mathbf{b}^{(\mathrm{A})}_1(\mathbf{s}_1^{(\mathrm{D})})=1$, where $\mathbf{0}_n$ is the n-dimensional zero-vector and $\mathbf{s}_1^{(\mathrm{D})}$ is given by the infrastructure configuration (see \S \ref{sec:system_model}).

\vspace{1mm}

\noindent\textbf{Best response strategies.} A defender strategy $\tilde{\pi}_{\mathrm{D}} \in \Pi_{\mathrm{D}}$ is called a \textit{best response} against $\pi_{\mathrm{A}}\in \Pi_{\mathrm{A}}$ if it \textit{maximizes} $J$ (\ref{eq:objective_fun}). Similarly, an attacker strategy $\tilde{\pi}_{\mathrm{A}}$ is called a best response against $\pi_{\mathrm{D}}$ if it \textit{minimizes} $J$ (\ref{eq:objective_fun}). Hence, the best response correspondences are
\begin{align}
  B_{\mathrm{D}}(\pi_{\mathrm{A}}) &\triangleq \argmax_{\pi_{\mathrm{D}} \in \Pi_{\mathrm{D}}}\mathbb{E}_{(\pi_{\mathrm{D}},\pi_{\mathrm{A}})}[J]\label{eq:br_defender}\\
  B_{\mathrm{A}}(\pi_{\mathrm{D}}) &\triangleq \argmin_{\pi_{\mathrm{A}} \in \Pi_{\mathrm{A}}} \mathbb{E}_{(\pi_{\mathrm{D}},\pi_{\mathrm{A}})}[J]\label{eq:br_attacker}
\end{align}

\vspace{1mm}

\noindent\textbf{Optimal strategies.} An optimal defender strategy $\pi_{\mathrm{D}}^{*}$ is a best response strategy against any attacker strategy that \textit{minimizes} $J$. Similarly, an optimal attacker strategy $\pi_{\mathrm{A}}^{*}$ is a best response against any defender strategy that \textit{maximizes} $J$. Hence, when both players follow optimal strategies, they play best response strategies against each other:
\begin{align}
  (\pi_{\mathrm{D}}^{*}, \pi_{\mathrm{A}}^{*}) \in B_{\mathrm{D}}(\pi_{\mathrm{A}}^{*}) \times B_{\mathrm{A}}(\pi_{\mathrm{D}}^{*})\label{eq:minmax_objective}
\end{align}
Since no player has an incentive to change its strategy, $(\pi_{\mathrm{D}}^{*},\pi_{\mathrm{A}}^{*})$ is a Nash equilibrium \cite[Eq. 1]{nash51}.

We know from game theory that $\Gamma$ has a mixed Nash equilibrium \cite{posg_equilibria_existence_finite_horizon,horak_thesis,po_posgs_horak_bosansky} and we know from Markov decision theory that $B_{\mathrm{D}}(\pi_{\mathrm{A}})$ and $B_{\mathrm{A}}(\pi_{\mathrm{D}})$ are non-empty \cite{krishnamurthy_2016,puterman}. Based on these standard results, we state the following theorem.
\begin{theorem}\label{thm:equilibrium}
$\quad$
  \begin{enumerate}[(A)]
  \item A game $\Gamma$ (\ref{def:game}) with instantiation described in \S \ref{sec:system_model} has a mixed Nash equilibrium.
  \item The best response correspondences (\ref{eq:br_defender})--(\ref{eq:br_attacker}) in $\Gamma$ with the instantiation described in \S \ref{sec:system_model} satisfy $|B_{\mathrm{D}}(\pi_{\mathrm{A}})|>0$ and $|B_{\mathrm{A}}(\pi_{\mathrm{D}})| > 0$ $\forall (\pi_{\mathrm{A}},\pi_{\mathrm{D}}) \in \Pi_{\mathrm{A}} \times \Pi_{\mathrm{D}}$.
  \end{enumerate}
\end{theorem}
\begin{proof}
  The statement in (A) follows from the following sufficient conditions: (\textit{i}) $\Gamma$ is stationary, finite, and zero-sum; (\textit{ii}) $\Gamma$ has public observations; and (\textit{iii}) $\gamma \in [0,1)$. Due to these conditions, the existence proofs in \cite[\S 3]{posg_equilibria_existence_finite_horizon}, \cite[Thm. 2.3]{horak_thesis}, and \cite[Thm. 1]{po_posgs_horak_bosansky} apply, which show that $\Gamma$ can be modeled as a finite strategic game, for which Nash's theorem applies \cite[Thm. 1]{nash51}. In the interest of space we do not restate the proof.

To prove (B), we note that obtaining a pair of best response strategies $(\tilde{\pi}_{\mathrm{D}} ,\tilde{\pi}_{\mathrm{A}}) \in B_{\mathrm{D}}(\pi_{\mathrm{A}})\times B_{\mathrm{A}}(\pi_{\mathrm{D}})$ for a given strategy pair $(\pi_{\mathrm{A}},\pi_{\mathrm{D}}) \in \Pi_{\mathrm{A}} \times \Pi_{\mathrm{D}}$ amounts to solving two finite and stationary \pomdps (Partially Observed Markov Decision Processes) with discounted utilities. It then follows from Markov decision theory that a pair of pure best response strategies $(\tilde{\pi}_{\mathrm{D}}, \tilde{\pi}_{\mathrm{A}})$ exists \cite[Thm. 6.2.7]{puterman} \cite[Thms. 7.6.1-7.6.2]{krishnamurthy_2016}. For the sake of brevity we do not restate the proof, which is based on Banach's fixed-point theorem \cite[Thm. 6, p. 160]{Banach1922}.
\end{proof}

\begin{table}
  \centering
  \begin{tabular}{ll} \toprule
    {\textit{Notation(s)}} & {\textit{Description}} \\ \midrule
    $\mathcal{G}, \mathcal{G}_{\mathbf{w}}$ & Infrastructure tree, subtree of $\mathbf{w}$\\
    $\mathcal{V}, \mathcal{E}$ & Sets of nodes and edges in $\mathcal{G}$\\
    $\mathcal{V}_{\mathbf{w}}, \mathcal{E}_{\mathbf{w}}$ & Sets of nodes and edges in $\mathcal{G}_{\mathbf{w}}$\\
    $\mathcal{Z}, \mathcal{W}$ & Sets of network zones and workflows\\
    $\mathcal{A}_{\mathrm{D}}, \mathcal{A}_{\mathrm{A}}(\mathbf{s}_t)$ & Defender and attacker action spaces at time $t$ \\
    $\mathcal{A}^{(\mathrm{V})}_{\mathrm{D}}, \mathcal{A}^{(\mathrm{V})}_{\mathrm{A}}(\mathbf{s}_t)$ & Action spaces per node at time $t$, $\mathcal{A}_{\mathrm{k}}=(\mathcal{A}^{(\mathrm{V})}_{\mathrm{k}})^{|\mathcal{V}|}$\\
    $\mathcal{O}^{(\mathrm{V})}$ & Observation space per node at time $t$, $\mathcal{O}=(\mathcal{O}^{(\mathrm{V})})^{|\mathcal{V}|}$\\
    $v_{i,t}^{(\mathrm{I})},v_{i,t}^{(\mathrm{Z})},v_{i,t}^{(\mathrm{R})}$ & Intrusion state and zone of $i \in \mathcal{V}$ at time $t$\\
    $v_{i,t}^{(\mathrm{R})}$ & Reconnaissance state of $i \in \mathcal{V}$ at time $t$\\
$V_{i,t}^{(\mathrm{I})},V_{i,t}^{(\mathrm{Z})},V_{i,t}^{(\mathrm{R})}$ & Random variables with realizations $v_{i,t}^{(\mathrm{I})},v_{i,t}^{(\mathrm{Z})},v_{i,t}^{(\mathrm{R})}$ \\
    $\Gamma, \mathcal{N}$ & \poposg (\ref{def:game}), set of players\\
    $\mathcal{S}_{\mathrm{D}}, \mathcal{S}_{\mathrm{A}}$ & Defender and attacker state spaces\\
    $\mathcal{S}\triangleq \mathcal{S}_{\mathrm{D}} \times \mathcal{S}_{\mathrm{A}}$ & State space\\
    $u, \mathcal{O}$ & Utility function and observation space\\
    $\mathbf{s}_t=(\mathbf{s}_t^{(\mathrm{D})},\mathbf{s}_t^{(\mathrm{A})})$ & State at time $t$\\
    $\mathbf{a}_t=(\mathbf{a}_t^{(\mathrm{D})},\mathbf{a}_t^{(\mathrm{A})})$ & Action at time $t$\\
    $\mathbf{o}_t, \mathbf{u}_t$ & Observation, utility  at time $t$\\
    $\mathbf{a}_t^{(\mathrm{k})},\mathbf{h}_t^{(\mathrm{k})}$ & Action and history of player $\mathrm{k}$ at time $t$\\
    $\mathcal{B}_{\mathrm{k}}, \mathbf{b}_t^{(\mathrm{k})}$ & Belief space and belief state of player $k$\\
    $\tilde{\pi}_{\mathrm{k}}, \tilde{\mathbf{a}}^{(\mathrm{k})}$ & Best response strategy and action of player $k$\\
    $\mathbf{S}_t, \mathbf{O}_t, \mathbf{A}_t$ & Random vectors with realizations $\mathbf{s}_t, \mathbf{o}_t, \mathbf{a}_t$\\
    $\mathbf{U}_t,\mathbf{B}^{(\mathrm{k})}_t,\mathbf{H}^{(\mathrm{k})}_t$ & Random vectors with realizations $\mathbf{u}_t,\mathbf{b}^{(\mathrm{k})}_t,\mathbf{h}^{(\mathrm{k})}_t$\\
    $\pi_{\mathrm{k}}, Z$ & Strategy of player $k$, observation distribution\\
    $u^{(\mathrm{w})}_{i,t}$ & Workflow utility of node $i$ at time $t$\\
    $\bot, \mathrm{an}(i)$ & Null action, set of $i$ and its ancestors in $\mathcal{G}$\\
   $\alpha_{i,t}$ & Active status of node $i$ at time $t$\\
    $f_{\mathrm{A}},f_{\mathrm{D}}$ & Attacker and defender transition functions\\
    $B_{\mathrm{k}}$ & Best response correspondence of player $\mathrm{k}$\\
    $c^{(\mathrm{I})}_{i,t}$ & Intrusion cost associated with node $i$ at time $t$\\
    $c^{(\mathrm{A})}$ & Action cost function\\
    \bottomrule\\
  \end{tabular}
  \caption{Notations.}\label{tab:notation}
\end{table}
\section{Decomposing the Intrusion Response Game}\label{sec:game_dec}
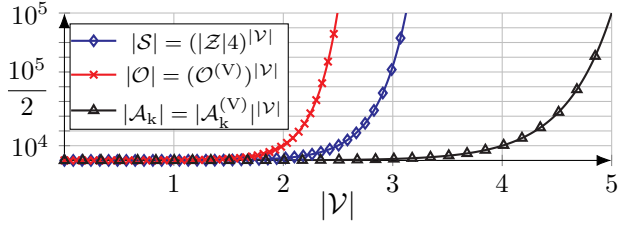
\begin{figure}
  \centering
  \scalebox{1}{
    \begin{tikzpicture}[
    dot/.style={
        draw=black,
        fill=blue!90,
        circle,
        minimum size=3pt,
        inner sep=0pt,
        solid,
    },
    ]
\tikzset{
        hatch distance/.store in=\hatchdistance,
        hatch distance=10pt,
        hatch thickness/.store in=\hatchthickness,
        hatch thickness=2pt
      }
\pgfdeclarepatternformonly[\hatchdistance,\hatchthickness]{flexible hatch}
    {\pgfqpoint{0pt}{0pt}}
    {\pgfqpoint{\hatchdistance}{\hatchdistance}}
    {\pgfpoint{\hatchdistance-1pt}{\hatchdistance-1pt}}%
    {
        \pgfsetcolor{\tikz@pattern@color}
        \pgfsetlinewidth{\hatchthickness}
        \pgfpathmoveto{\pgfqpoint{0pt}{0pt}}
        \pgfpathlineto{\pgfqpoint{\hatchdistance}{\hatchdistance}}
        \pgfusepath{stroke}
      }
\node[scale=1] (kth_cr) at (0,2.15)
{
  \begin{tikzpicture}[declare function={sigma(\x)=1/(1+exp(-\x));
sigmap(\x)=sigma(\x)*(1-sigma(\x));}]
    \begin{axis}[
        xmin=0,
        xmax=5,
        ymin=0,
        ymax=100000,
        width = \columnwidth,
        height = 0.4\columnwidth,
        axis lines=center,
        xtick={0,1,2,3,4,5},
        ytick={0,10^4, 2*10^4, 3*10^4, 4*10^4, 5*10^4, 6*10^4, 7*10^4, 8*10^4, 9*10^4, 10^5},
        grid = both,
        major grid style = {lightgray},
        minor grid style = {lightgray!25},
        yticklabels={$$, $10^4$, $$, $$, $$, $\displaystyle\frac{10^5}{2}$, $$, $$, $$, $$, $10^5$},
        scaled y ticks=false,
        yticklabel style={
        /pgf/number format/fixed,
        /pgf/number format/precision=5
        },
        xlabel style={below right},
        ylabel style={above left},
        axis line style={-{Latex[length=2mm]}},
        smooth,
        legend style={at={(0.41,0.93)}},
        legend columns=1,
        legend style={
            inner sep=0pt,
            font=\footnotesize,
            /tikz/column 2/.style={
                column sep=5pt,
              }
              }
        ]
\addplot[Blue,mark=diamond, mark repeat=2, samples=100,smooth, name path=l1, thick, domain=0:4, restrict y to domain=0:1000000]   (x,{10^x*2^(2*x)});

\addplot[Red,mark=x,mark repeat=4, samples=300,smooth, name path=l1, thick, domain=0:4, restrict y to domain=0:1000000]   (x,{(100+1)^x});

\addplot[Black,mark=triangle,mark repeat=10, samples=300,smooth, name path=l1, thick, domain=0:5, restrict y to domain=0:1000000]   (x,{(10)^x});

\legend{$|\mathcal{S}|=(|\mathcal{Z}|4)^{|\mathcal{V}|}$, $|\mathcal{O}|=(\mathcal{O}^{(\mathrm{V})})^{|\mathcal{V}|}$, $|\mathcal{A}_{\mathrm{k}}|=|\mathcal{A}^{(\mathrm{V})}_{\mathrm{k}}|^{|\mathcal{V}|}$}
\end{axis}
%
\node[inner sep=0pt,align=center, scale=1.1, rotate=0, opacity=1] (obs) at (3.7,-0.6)
{
  $|\mathcal{V}|$
};
\end{tikzpicture}
};

  \end{tikzpicture}
  }
  \caption{Growth of $|\mathcal{S}|$, $|\mathcal{O}|$, and $|\mathcal{A}_{\mathrm{k}}|$ in function of the number of nodes $|\mathcal{V}|$, where $\mathrm{k} \in \{\mathrm{D}, \mathrm{A}\}$; the curves are computed using $|\mathcal{Z}|=10$, $|\mathcal{O}^{(\mathrm{V})}|=100$, and $|\mathcal{A}_{\mathrm{D}}^{(\mathrm{V})}|=|\mathcal{A}_{\mathrm{A}}^{(\mathrm{V})}|=10$.}
  \label{fig:spaces_scale}
\end{figure}

\begin{figure*}
\captionsetup[subfigure]{justification=centering}
  \centering
    \begin{subfigure}[t]{0.25\textwidth}
  \centering
  \scalebox{0.61}{
          \begin{tikzpicture}[fill=white, >=stealth,
    node distance=3cm,
    database/.style={
      cylinder,
      cylinder uses custom fill,
      shape border rotate=90,
      aspect=0.25,
      draw}]

    \tikzset{
node distance = 9em and 4em,
sloped,
   box/.style = {%
    shape=rectangle,
    rounded corners,
    draw=blue!40,
    fill=blue!15,
    align=center,
    font=\fontsize{12}{12}\selectfont},
 arrow/.style = {%
    line width=0.1mm,
    -{Triangle[length=5mm,width=2mm]},
    shorten >=1mm, shorten <=1mm,
    font=\fontsize{8}{8}\selectfont},
}
\node[scale=1] (node2) at (1.2,4.4)
{
  \begin{tikzpicture}
\node[scale=1] (b1) at (0,0)
{
  \begin{tikzpicture}
\draw[-, color=black] (0,0) to (1.2,0) to (1.2,1) to (0, 1) to (0,0);
    \end{tikzpicture}
  };
\node[inner sep=0pt,align=center, scale=1] (time) at (0.1,0.0)
{
  $\pi_{\mathrm{k}}^{(\mathbf{w}_1)}$
};

\node[scale=1] (b1) at (0,-1.25)
{
  \begin{tikzpicture}
\draw[-, color=black] (0,0) to (1.2,0) to (1.2,1) to (0, 1) to (0,0);
    \end{tikzpicture}
  };
  \node[inner sep=0pt,align=center, scale=1] (time) at (0.1,-1.25)
{
  $\pi_{\mathrm{k}}^{(\mathbf{w}_2)}$
};

\node[scale=1] (b1) at (0,-3)
{
  \begin{tikzpicture}
\draw[-, color=black] (0,0) to (1.2,0) to (1.2,1) to (0, 1) to (0,0);
    \end{tikzpicture}
  };
  \node[inner sep=0pt,align=center, scale=1] (time) at (0.1,-3)
{
  $\pi_{\mathrm{k}}^{(\mathbf{w}_{|\mathcal{W}|})}$
};

\draw[->, color=black, line width=0.4mm] (-1.5,0) to (-0.62, 0);
\draw[->, color=black, line width=0.4mm] (-1.5,-1.25) to (-0.62, -1.25);
\draw[->, color=black, line width=0.4mm] (-1.5,-3) to (-0.62, -3);
\node[inner sep=0pt,align=center, scale=1] (time) at (-1.1,0.35)
{
$\mathbf{h}^{(\mathrm{k})}_{\mathbf{w}_1,t}$
};
\node[inner sep=0pt,align=center, scale=1] (time) at (-1.1,-0.95)
{
$\mathbf{h}^{(\mathrm{k})}_{\mathbf{w}_2,t}$
};
\node[inner sep=0pt,align=center, scale=1] (time) at (-1.1,-2.6)
{
$\mathbf{h}^{(\mathrm{k})}_{\mathbf{w}_{|\mathcal{W}|},t}$
};
\node[inner sep=0pt,align=center, scale=1.7] (time) at (0.1,-1.95)
{
$\vdots$
};

\node[inner sep=0pt,align=center, scale=1.7] (oplus) at (1.8,-1.25)
{
$\oplus$
};
\draw[->, color=black, line width=0.4mm] (0.6,0) to (1.7,0) to (1.7, -1.05);
\draw[->, color=black, line width=0.4mm] (0.6,-1.25) to (oplus);
\draw[->, color=black, line width=0.4mm] (0.6,-3) to (1.7,-3) to (1.7, -1.45);

\draw[->, color=black, line width=0.4mm] (1.88,-1.25) to (2.6,-1.25);

\node[inner sep=0pt,align=center, scale=1] (oplus) at (1.1,0.3)
{
$\mathbf{a}^{(\mathrm{k})}_{\mathbf{w}_1,t}$
};
\node[inner sep=0pt,align=center, scale=1] (oplus) at (1.1,-0.91)
{
$\mathbf{a}^{(\mathrm{k})}_{\mathbf{w}_2,t}$
};
\node[inner sep=0pt,align=center, scale=1] (oplus) at (1.225,-2.65)
{
$\mathbf{a}^{(\mathrm{k})}_{\mathbf{w}_{|\mathcal{W}|},t}$
};

\node[inner sep=0pt,align=center, scale=1] (time) at (2.25,-0.91)
{
$\mathbf{a}^{(\mathrm{k})}_{t}$
};
\end{tikzpicture}
};
\end{tikzpicture}
  }
  \caption{Theorem \ref{thm:decomp}.A}
  \label{fig:workflow_decomposition}
    \end{subfigure}
    \hfill
    \begin{subfigure}[t]{0.4\textwidth}
  \centering
  \scalebox{0.61}{
          \begin{tikzpicture}[fill=white, >=stealth,
    node distance=3cm,
    database/.style={
      cylinder,
      cylinder uses custom fill,
      shape border rotate=90,
      aspect=0.25,
      draw}]

    \tikzset{
node distance = 9em and 4em,
sloped,
   box/.style = {%
    shape=rectangle,
    rounded corners,
    draw=blue!40,
    fill=blue!15,
    align=center,
    font=\fontsize{12}{12}\selectfont},
 arrow/.style = {%
    line width=0.1mm,
    -{Triangle[length=5mm,width=2mm]},
    shorten >=1mm, shorten <=1mm,
    font=\fontsize{8}{8}\selectfont},
}

\node[scale=1] (node1) at (0,0)
{
\begin{tikzpicture}

\node[scale=1] (reach1) at (6,0)
{
  \begin{tikzpicture}
\draw[-, color=black] (0,0) to (1.5,0) to (1.5,1) to (0, 1) to (0,0);
\node[inner sep=0pt,align=center, scale=1.1] (time) at (0.85,0.5)
{
  $\pi^{(1)}_{\mathrm{k}}$
};
    \end{tikzpicture}
  };
\node[draw,circle, minimum width=2mm, scale=0.6, fill=black](pijoin01) at (4.4,0) {};
\draw[->, color=black, line width=0.4mm] (pijoin01) to (5.25,0);
\node[inner sep=0pt,align=center, scale=1] (time) at (2.9,0.3)
{
$\mathbf{h}^{(\mathrm{k})}_{1,t}$
};
\node[inner sep=0pt,align=center, scale=1.7] (oplus) at (3.75,0)
{
$\oplus$
};

\draw[-, color=black, line width=0.4mm] (3.85,0) to (pijoin01);
\draw[->, color=black, line width=0.4mm] (2.33,0) to (3.45, 0);

\end{tikzpicture}
};

\node[scale=1] (node1) at (0,-1.3)
{
\begin{tikzpicture}
\node[scale=1] (reach1) at (6,0)
{
  \begin{tikzpicture}
\draw[-, color=black] (0,0) to (1.5,0) to (1.5,1) to (0, 1) to (0,0);
\node[inner sep=0pt,align=center, scale=1.1] (time) at (0.85,0.5)
{
  $\pi^{(2)}_{\mathrm{k}}$
};
    \end{tikzpicture}
  };
\node[draw,circle, minimum width=2mm, scale=0.6, fill=black](pijoin01) at (4.4,0) {};
\draw[->, color=black, line width=0.4mm] (pijoin01) to (5.25,0);
\node[inner sep=0pt,align=center, scale=1] (time) at (2.9,0.3)
{
$\mathbf{h}^{(\mathrm{k})}_{2,t}$
};

\node[inner sep=0pt,align=center, scale=1.7] (oplus) at (3.75,0)
{
$\oplus$
};

\draw[-, color=black, line width=0.4mm] (3.85,0) to (pijoin01);
\draw[->, color=black, line width=0.4mm] (2.33,0) to (3.45, 0);
\end{tikzpicture}
};

\node[scale=1] (node1) at (0,-3.05)
{
\begin{tikzpicture}

\node[scale=1] (reach1) at (6,0)
{
  \begin{tikzpicture}
\draw[-, color=black] (0,0) to (1.5,0) to (1.5,1) to (0, 1) to (0,0);
    \end{tikzpicture}
  };
\node[inner sep=0pt,align=center, scale=1.1] (time) at (6.1,0)
{
$\pi^{(|\mathcal{V}_{\mathrm{w}}|)}_{\mathrm{k}}$
};
\node[inner sep=0pt,align=center, scale=1] (time) at (2.9,0.36)
{
$\mathbf{h}^{(\mathrm{k})}_{|\mathcal{V}_{\mathbf{w}}|,t}$
};

\node[inner sep=0pt,align=center, scale=1.7] (oplus) at (3.75,0)
{
$\oplus$
};

\draw[->, color=black, line width=0.4mm] (3.85,0) to (5.25,0);
\draw[->, color=black, line width=0.4mm] (2.33,0) to (3.45, 0);
\end{tikzpicture}
};

\node[inner sep=0pt,align=center, scale=1.7] (time) at (1.5,-2)
{
$\vdots$
};

\node[inner sep=0pt,align=center, scale=1.7] (join11) at (3.79,-1.3)
{
$\oplus$
};
\draw[->, color=black, line width=0.4mm] (2.16, -1.3) to (join11);

\node[inner sep=0pt,align=center, scale=1] (oplus) at (2.85,-1)
{
$(\mathbf{a}^{(\mathrm{k})}_{2,t})$
};

\draw[->, color=black, line width=0.4mm] (3.87,-1.3) to (4.7, -1.3);

\node[inner sep=0pt,align=center, scale=1] (oplus) at (4.25,-1)
{
$\mathbf{a}^{(\mathrm{k})}_{\mathbf{w},t}$
};

\node[inner sep=0pt,align=center, scale=1] (oplus) at (3.05,-2.7)
{
$(\mathbf{a}^{(\mathrm{k})}_{|\mathcal{V}_{\mathbf{w}}|,t})$
};

\draw[->, color=black, line width=0.4mm] (2.16, -3.05) to (3.7, -3.05) to (3.7, -1.5);

\draw[->, color=black, line width=0.4mm] (2.16, 0) to (3.7, 0) to (3.7, -1.1);

\node[inner sep=0pt,align=center, scale=1] (oplus) at (2.85,0.3)
{
$(\mathbf{a}^{(\mathrm{k})}_{1,t})$
};
\draw[->, color=black, line width=0.4mm] (-0.19, 0) to (-0.19, -0.55) to (-0.9, -0.55) to (-0.9, -1.1);
\draw[->, color=black, line width=0.4mm] (-0.19, -1.3) to (-0.19, -1.85) to (-0.9, -1.85) to (-0.9, -2.85);
\draw[->, color=black, line width=0.4mm] (-0.9, 0.75) to (-0.9, 0.2);

\node[inner sep=0pt,align=center, scale=1] (oplus) at (-0.2,0.5)
{
$|\mathrm{an}(0)|$
};

\node[inner sep=0pt,align=center, scale=1] (oplus) at (-0.2,-0.9)
{
$|\mathrm{an}(1)|$
};

\node[inner sep=0pt,align=center, scale=1] (oplus) at (-0.05,-2.4)
{
$|\mathrm{an}(|\mathcal{V}_{\mathbf{w}}|)|$
};

\end{tikzpicture}
  }
  \caption{Theorem \ref{thm:decomp}.B}  \label{fig:decomposition4}
    \end{subfigure}
    \hfill
    \begin{subfigure}[t]{0.25\textwidth}
  \centering
  \scalebox{0.28}{
    \begin{tikzpicture}[declare function={sigma(\x)=1/(1+exp(-\x));
    sigmap(\x)=sigma(\x)*(1-sigma(\x));}]

\node[scale=0.8] (kth_cr) at (0,2.15)
{
  \begin{tikzpicture}
\draw[-, black, thick, line width=0.9mm, name path=simplex] (0,0) to (10,0) to (5,10) to (0,0);
\draw[-, black, thick, line width=0.9mm, name path=simplex2] (10,0) to (0,0) to (5,10);

\node[draw,circle, fill=black, scale=1](point) at (7.57,4.85) {};
\draw[-, black, dashed, thick, line width=0.5mm] (point) to (0, 0);

\node[draw,circle, fill=black, scale=1](point1) at (8.08,3.8) {};
\draw[-, black, dashed, thick, line width=0.5mm] (point1) to (0, 0);

\node[draw,circle, fill=black, scale=1](point2) at (8.63,2.7) {};
\draw[-, black, dashed, thick, line width=0.5mm] (point2) to (0, 0);

\node[draw,circle, fill=black, scale=1](point3) at (9.17,1.6) {};
\draw[-, black, dashed, thick, line width=0.5mm] (point3) to (0, 0);

\node[draw,circle, fill=black, scale=1](point4) at (9.68,0.6) {};
\draw[-, black, dashed, thick, line width=0.5mm] (point4) to (0, 0);

\node[draw,circle, fill=black, scale=1](point5) at (7.05,5.9) {};
\draw[-, black, dashed, thick, line width=0.5mm] (point5) to (0, 0);

\node[draw,circle, fill=black, scale=1](point6) at (6.502,6.9) {};
\draw[-, black, dashed, thick, line width=0.5mm] (point6) to (0, 0);

\node[draw,circle, fill=black, scale=1](point7) at (6,7.94) {};
\draw[-, black, dashed, thick, line width=0.5mm] (point7) to (0, 0);

\node[draw,circle, fill=black, scale=1](point8) at (5.5,8.94) {};
\draw[-, black, dashed, thick, line width=0.5mm] (point8) to (0, 0);


\draw [decorate,decoration={brace,amplitude=9pt,mirror,raise=4pt},yshift=0pt,rotate=118, line width=0.5mm]
(-5,-10.45) -- (6,-10.45) node [black,midway,xshift=0.2cm] {};

\node[inner sep=0pt,align=center, scale=2.2, rotate=0, opacity=1] (obs) at (12.9,5.7)
{
  sub-simplex $\mathcal{B}_{\mathrm{D},\mathbf{e}_1}^{(i)}$\\
  joining $\mathbf{e}_2$ and $\mathbf{e}_3$
};

\node[inner sep=0pt,align=center, scale=2.2, rotate=0, opacity=1] (obs) at (12.6,3)
{
  Stopping set\\
  $\mathscr{S}^{(i)}_{\mathrm{D}}$
};

\node[inner sep=0pt,align=center, scale=2, rotate=0, opacity=1] (obs) at (8.56,5.15)
{
$\widehat{\mathbf{b}}^{(\mathrm{D})}_5$
};

\node[inner sep=0pt,align=center, scale=2, rotate=0, opacity=1] (obs) at (8.15,6.15)
{
$\widehat{\mathbf{b}}^{(\mathrm{D})}_4$
};

\node[inner sep=0pt,align=center, scale=2, rotate=0, opacity=1] (obs) at (7.55,7.15)
{
$\widehat{\mathbf{b}}^{(\mathrm{D})}_3$
};

\node[inner sep=0pt,align=center, scale=2, rotate=0, opacity=1] (obs) at (7.05,8.15)
{
$\widehat{\mathbf{b}}^{(\mathrm{D})}_2$
};

\node[inner sep=0pt,align=center, scale=2, rotate=0, opacity=1] (obs) at (6.55,9.15)
{
$\widehat{\mathbf{b}}^{(\mathrm{D})}_1$
};

\node[inner sep=0pt,align=center, scale=2, rotate=0, opacity=1] (obs) at (9.1,4.05)
{
$\widehat{\mathbf{b}}^{(\mathrm{D})}_6$
};

\node[inner sep=0pt,align=center, scale=2, rotate=0, opacity=1] (obs) at (9.7,2.95)
{
$\widehat{\mathbf{b}}^{(\mathrm{D})}_7$
};

\node[inner sep=0pt,align=center, scale=2, rotate=0, opacity=1] (obs) at (10.25,1.85)
{
$\widehat{\mathbf{b}}^{(\mathrm{D})}_8$
};

\node[inner sep=0pt,align=center, scale=2, rotate=0, opacity=1] (obs) at (10.7,0.85)
{
$\widehat{\mathbf{b}}^{(\mathrm{D})}_9$
};

\node[inner sep=0pt,align=center, scale=2.2, rotate=0, opacity=1] (obs) at (-0.6,5)
{
  Continuation set\\
  $\mathscr{C}^{(i)}_{\mathrm{D}}$
};

\node[inner sep=0pt,align=center, scale=2.2, rotate=0, opacity=1] (obs) at (-1.8,2.1)
{
  Belief space\\
  $\mathcal{B}^{(i)}_{\mathrm{D}}$
};

\node[inner sep=0pt,align=center, scale=2.2, rotate=0, opacity=1] (obs) at (2.1,9.84)
{
$\mathcal{L}(\mathbf{e}_1, \widehat{\mathbf{b}}^{(\mathrm{D})}_3)$
};
\node[inner sep=0pt,align=center, scale=2.2, rotate=0, opacity=1] (obs) at (0.53,7.15)
{
  Switching curve\\
  $\Upsilon$
};

\node[inner sep=0pt,align=center, scale=2.2, rotate=0, opacity=1] (obs) at (5.6,-1)
{
  Threshold belief state $\alpha_{\widehat{\mathbf{b}}_8}$
};

\node[draw,circle, fill=Yellow, scale=1.05](threshold9) at (6.8,0.44) {};
\node[draw,circle, fill=Yellow, scale=1.05](threshold8) at (6.5,1.25) {};
\node[draw,circle, fill=Yellow, scale=1.05](threshold7) at (6,2.15) {};
\node[draw,circle, fill=Yellow, scale=1.05](threshold6) at (5.7,2.9) {};
\node[draw,circle, fill=Yellow, scale=1.05](threshold5) at (5.4,3.75) {};
\node[draw,circle, fill=Yellow, scale=1.05](threshold4) at (5,4.4) {};
\node[draw,circle, fill=Yellow, scale=1.05](threshold3) at (4.65,5.1) {};
\node[draw,circle, fill=Yellow, scale=1.05](threshold2) at (4.3,5.75) {};
\node[draw,circle, fill=Yellow, scale=1.05](threshold1) at (3.9,6.35) {};
\draw[line width=0.4mm, name path=switching] (3.4, 6.8) to (threshold1) to (threshold2) to (threshold3) to (threshold4) to (threshold5) to (threshold6) to (threshold7) to (threshold8) to (threshold9) to (7.2,0);

\node[inner sep=0pt,align=center, scale=2.2, rotate=0, opacity=1] (obs) at (-1,-0.5)
{
  $\mathbf{e}_1$\\
  $(1,0,0)$
};
\node[inner sep=0pt,align=center, scale=2.2, rotate=0, opacity=1] (obs) at (11,-0.5)
{
  $\mathbf{e}_2$\\
  $(0,1,0)$
};

\node[inner sep=0pt,align=center, scale=2.2, rotate=0, opacity=1] (obs) at (5,10.85)
{
  $\mathbf{e}_3$\\
  $(0,0,1)$
};

    \path[
pattern=grid,
        intersection segments={
                of=switching and simplex2,
                sequence={R2--L2}
              }];

\draw[line width=0.4mm, -{Latex[length=5mm]},color=Red] (8.2,-0.55) to (threshold8);
\draw[line width=0.4mm, -{Latex[length=5mm]},color=Red] (2.4,9.3) to (5.87, 6.4);
\draw[line width=0.4mm, -{Latex[length=5mm]},color=Red] (1.5,7) to (3.3, 6.65);
\draw[line width=0.4mm, -{Latex[length=5mm]},color=Red] (0,4.7) to (2.3, 3.4);
\draw[line width=0.4mm, -{Latex[length=5mm]},color=Red] (-0.9,2.1) to (0.4, 0.75);
\draw[line width=0.4mm, -{Latex[length=5mm]},color=Red] (11.5,2.65) to (8, 2);
  \end{tikzpicture}
};

\end{tikzpicture}
  }
  \caption{Theorem \ref{thm:decomp}.C}\label{fig:unit_simplex_2}
\end{subfigure}
\caption{Illustrations of Thm. \ref{thm:decomp}; arrows indicate inputs and outputs; $\oplus$ denotes vector concatenation; $k \in \{\mathrm{D}, \mathrm{A}\}$; $\mathbf{h}^{(\mathrm{k})}_{\mathbf{w},t}\triangleq (\mathbf{h}^{(\mathrm{k})}_{j,t})_{j \in \mathcal{V}_{\mathbf{w}}}$; and $\mathbf{a}^{(\mathrm{k})}_{\mathbf{w},t}\triangleq (\mathbf{a}^{(\mathrm{k})}_{j,t})_{j \in \mathcal{V}_{\mathbf{w}}}$; (a) illustrates that a game strategy $\pi_{\mathrm{k}}$ decomposes into $|\mathcal{W}|$ independent substrategies; (b) illustrates that a workflow strategy $\pi^{(\mathbf{w})}_{\mathrm{k}}$ for $\mathbf{w} \in \mathcal{W}$ decomposes into substrategies $(\pi_{\mathrm{k}}^{(i)})_{i\in\mathcal{V}_{\mathbf{w}}}$ with optimal substructure; (c) provides a geometric illustration of the proof of Thm. \ref{thm:decomp}.C, showing a switching curve that partitions the defender's belief space of a node $i \in \mathcal{V}$.}
    \label{fig:theorem_illustrations}
\end{figure*}
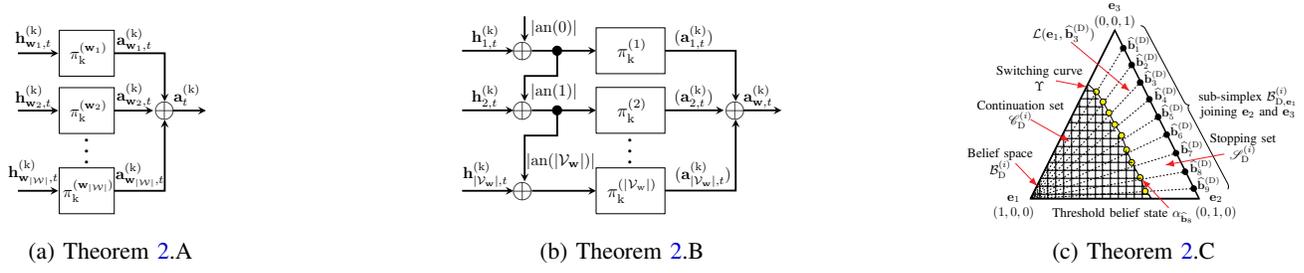

In this section we present the main contribution of the paper. We show how the game $\Gamma$ (\ref{def:game}) with the instantiation described in \S \ref{sec:system_model} can be recursively decomposed into subgames with optimal substructure \cite[Ch. 15]{cormen01introduction}, which means that a best response (\ref{eq:br_defender})--(\ref{eq:br_attacker}) of the original game can be obtained from best responses of the subgames. We further show that best responses of the subgames can be computed in parallel and that the space complexity of a subgame is independent of the number of nodes $|\mathcal{V}|$. Note that the space complexity of the original game increases exponentially with $|\mathcal{V}|$ (see Fig. \ref{fig:spaces_scale}).
\begin{theorem}[Decomposition theorem]\label{thm:decomp}
$\quad$
\begin{enumerate}[(A)]
\item A game $\Gamma$ (\ref{def:game}) with the instantiation described in \S \ref{sec:system_model} can be decomposed into independent workflow subgames $\Gamma^{(\mathbf{w}_1)},\hdots,\Gamma^{(\mathbf{w}_{|\mathcal{W}|})}$. Due to their independence, the subgames have optimal substructure.
\item Each subgame $\Gamma^{(\mathbf{w})}$ can be further decomposed into node subgames $(\Gamma^{(i)})_{i \in \mathcal{V}_{\mathbf{w}}}$ with optimal substructure and space complexities independent of $|\mathcal{V}|$.
\item For each subgame $\Gamma^{(i)}$, a best response strategy for the defender can be characterized by switching curves, under the assumption that the observation distributions $Z_{\mathbf{O}_1\mid \mathbf{s}^{(\mathrm{A})}},\hdots, Z_{\mathbf{O}_{|\mathcal{V}|}\mid \mathbf{s}^{(\mathrm{A})}}$ (\ref{eq:obs_1}) are totally positive of order 2 (i.e., \tpp \cite[Def. 10.2.1]{krishnamurthy_2016}).
\end{enumerate}
\end{theorem}
Statements A and B express that $\Gamma$ decomposes into simpler subgames, which consequently can be solved in parallel (see Fig. \ref{fig:theorem_illustrations}). This decomposition implies that the largest game that is tractable on a given compute platform scales linearly with the number of processors. Further, statement C says that a best response strategy for the defender in each subgame can be characterized by switching curves, which can be estimated efficiently.

In the following sections we provide proofs of Thm. \ref{thm:decomp}.A--C. The requisite notations are given in Table \ref{tab:notation}.

\subsection{Proof of Theorem \ref{thm:decomp}.A}
Following the instantiation of $\Gamma$ described in \S \ref{sec:system_model}, the state, observation, and action spaces factorize as
\begin{align}
\mathcal{S} = (\mathcal{Z}\times \{0,1\}^{2})^{|\mathcal{V}|}, \mathcal{O} = (\mathcal{O}^{(\mathrm{V})})^{|\mathcal{V}|}, \mathcal{A}_{\mathrm{k}} = (\mathcal{A}_{\mathrm{k}}^{(\mathrm{V})})^{|\mathcal{V}|}\label{eq:factorization:a}
\end{align}
for player $\mathrm{k} \in \{\mathrm{D}, \mathrm{A}\}$, where $\mathcal{O}^{(\mathrm{V})}$, $\mathcal{A}_{\mathrm{D}}^{(\mathrm{V})}$, and $\mathcal{A}_{\mathrm{A}}^{(\mathrm{V})}$ denote the local observation and action spaces for each node.

Since each node belongs to exactly one workflow, (\ref{eq:factorization:a}) implies that $\Gamma$ can be decomposed into subgames $\Gamma^{(\mathbf{w}_1)},\hdots,\Gamma^{(\mathbf{w}_{|\mathcal{W}|})}$. To show that the subgames are independent, it suffices to show that they are observation-independent, transition-independent, and utility-independent \cite[Defs. 32,33,35]{decomposed_mdps}.

From (\ref{eq:obs_1}) we have
\begin{align}
Z\big(\mathbf{o}_{i,t+1} \mid \mathbf{s}^{(\mathrm{D})}_{t+1},\mathbf{s}^{(\mathrm{A})}_{t+1}\big) &= Z\big(\mathbf{o}_{i,t+1} \mid \mathbf{s}^{(\mathrm{D})}_{i,t+1},\mathbf{s}^{(\mathrm{A})}_{i,t+1}\big)
\end{align}
for all $\mathbf{o}_{i,t+1} \in \mathcal{O}$, $\mathbf{s}_{t+1}\in \mathcal{S}$, and $t \geq 1$, which implies observation independence across nodes $i \in \mathcal{V}$ and therefore across workflows \cite[Def. 33]{decomposed_mdps}.

From the definitions in \S \ref{sec:system_model} and (\ref{eq:transition_atc})--(\ref{eq:transition_def}) we have
\begin{align}
f_{\mathrm{D}}(\mathbf{s}^{(\mathrm{D})}_{i,t+1} | \mathbf{s}^{(\mathrm{D})}_{t},\mathbf{a}^{(\mathrm{D})}_{t}) &= f_{\mathrm{D}}(\mathbf{s}^{(\mathrm{D})}_{i,t+1} | \mathbf{s}^{(\mathrm{D})}_{i,t},\mathbf{a}^{(\mathrm{D})}_{i,t})\nonumber\\
f_{\mathrm{A}}(\mathbf{s}^{(\mathrm{A})}_{i,t+1} | \mathbf{s}^{(\mathrm{A})}_{t},\mathbf{a}^{(\mathrm{A})}_{t},\mathbf{a}^{(\mathrm{D})}_{t}) &= f_{\mathrm{A}}(\mathbf{s}^{(\mathrm{A})}_{i,t+1} | \mathbf{s}^{(\mathrm{A})}_{i,t},\mathbf{a}^{(\mathrm{A})}_{i,t},\mathbf{a}^{(\mathrm{D})}_{i,t})\nonumber
\end{align}
for all $\mathbf{s}_{i,t} \in \mathcal{S},\mathbf{a}_{i,t} \in \mathcal{A}, i \in \mathcal{V}$, and $t \geq 1$, which implies transition independence across nodes $i \in \mathcal{V}$ and therefore across workflows \cite[Def. 32]{decomposed_mdps}.

Following (\ref{eq:objective_fun}) and the definition of $u_{i,t}^{(\mathrm{W})}$ (see \S \ref{sec:ir_problem}) we can rewrite $u(\mathbf{s}_t,\mathbf{a}_t^{(\mathrm{D})})$ as
\begin{align}
  u(\mathbf{s}_t,\mathbf{a}_t^{(\mathrm{D})})&=\sum_{\mathbf{w}\in \mathcal{W}}\overbrace{\sum_{i\in \mathcal{V}_{\mathbf{w}}}\eta u_{i,t}^{(\mathrm{W})}-c_{i,t}^{(\mathrm{I})}(\mathbf{a}^{(\mathrm{D})}_{i,t},v^{(\mathrm{I})}_{i,t})}^{\triangleq u_{\mathbf{w}}}\nonumber\\
                                             &=\sum_{\mathbf{w}\in \mathcal{W}} u_{\mathbf{w}}\Big(\big(\mathbf{s}_{i,t},\mathbf{a}^{(\mathrm{D})}_{i,t}\big)_{i \in \mathcal{V}_{\mathbf{w}}}\Big)\label{eq:workflow_util}
\end{align}
The final expression in (\ref{eq:workflow_util}) is a sum of workflow utility functions, each of which depends only on the states and actions of one workflow. Hence, $\Gamma^{(\mathbf{w}_1)},\hdots,\Gamma^{(\mathbf{w}_{|\mathcal{W}|})}$ are utility independent \cite[Def. 35]{decomposed_mdps}. \qed
\subsection{Proof of Theorem \ref{thm:decomp}.B}\label{proof:thm_2_b}
Our goal is to show that a workflow subgame $\Gamma^{(\mathbf{w})}$ decomposes into node-level subgames with optimal substructure. That is, we aim to show that a best response in $\Gamma^{(\mathbf{w})}$ can be constructed from best responses of the subgames.

Following the description in \S \ref{sec:system_model}, we know that the nodes in a workflow are connected in a tree and that the utility generated by a node $i$ depends on the number of active nodes in the subtree rooted at $i$. Taking into account this tree structure and the definition of the utility function, we decompose $\Gamma^{(\mathbf{w})}$ into node subgames $(\Gamma^{(i)})_{i \in \mathcal{V}_{\mathbf{w}}}$ where each subgame depends only on the local state and action of a single node. It follows from (\ref{eq:factorization:a}) that this decomposition is feasible and that the space complexity of a subgame is independent of $|\mathcal{V}|$. Further, we know from Thm. \ref{thm:decomp}.A that the subgames are transition-independent and observation-independent but utility-dependent. To prove optimal substructure it therefore suffices to show that it is possible to redefine the utility functions for the subgames such that at each time $t$, the best response action in $\Gamma^{(\mathbf{w})}$ for any node $i$ is also a best response in $\Gamma^{(i)}$ and vice versa. For the sake of brevity we give the proof for the defender only. The proof for the attacker is analogous. In this proof, for better readability, we omit the constants $\gamma,\eta$ and use the shorthand notations $\mathbf{s}_{\mathbf{w},t}^{(\mathrm{D})} \triangleq (\mathbf{s}^{(\mathrm{D})}_{j,t})_{j \in \mathcal{V}_{\mathbf{w}}}$, $\mathbf{b}_{\mathbf{w},t}^{(\mathrm{D})} \triangleq (\mathbf{b}^{(\mathrm{D})}_{j,t})_{j \in \mathcal{V}_{\mathbf{w}}}$, $\mathscr{V}\triangleq \mathscr{V}_{\mathrm{D},\pi_{\mathrm{A}}}^{*}$, and $\tau \in \argmin_{k>t}\mathbf{a}^{(\mathrm{D})}_k\neq \bot$, where $\mathscr{V}$ is the value function \cite[Thm. 7.4.1]{krishnamurthy_2016}. Further, we use $\mathrm{an}(i)$ to denote the set of node $i$ and its ancestors in the infrastructure graph $\mathcal{G}$.

From Bellman's optimality equation \cite[Eq. 1]{bellman1957markovian}, a best response action for node $i$ at time $t$ in $\Gamma^{(\mathbf{w})}$ against an attacker strategy $\pi_{\mathrm{A}}$ is given by
\begin{align}
&\argmax_{\mathbf{a}_{i,t}^{(\mathrm{D})}\in \mathcal{A}^{(\mathrm{V})}_{\mathrm{D}}}\Biggl[\underset{\pi_{\mathrm{A}}}{\mathbb{E}}\biggl[\mathbf{U}_{t} + \mathscr{V}(\mathbf{S}^{(\mathrm{D})}_{t+1}, \mathbf{B}^{(\mathrm{D})}_{t+1}) \Bigl\vert\mathbf{s}^{(\mathrm{D})}_t, \mathbf{b}^{(\mathrm{D})}_t, \mathbf{a}_{i,t}^{(\mathrm{D})}\biggr]\Biggr]\nonumber\\
&\numeq{a}\argmax_{\mathbf{a}_{i,t}^{(\mathrm{D})} \in \mathcal{A}^{(\mathrm{V})}_{\mathrm{D}}}\Biggl[\underset{\pi_{\mathrm{A}}}{\mathbb{E}}\biggl[-c^{(\mathrm{I})}_{i,t} + \mathscr{V}(\mathbf{S}^{(\mathrm{D})}_{t+1}, \mathbf{B}^{(\mathrm{D})}_{t+1}) \Bigl\vert\mathbf{s}^{(\mathrm{D})}_t, \mathbf{b}^{(\mathrm{D})}_t, \mathbf{a}_{i,t}^{(\mathrm{D})}\biggr]\Biggr]\nonumber\\
&\numeq{b}\argmax_{\mathbf{a}_{i,t}^{(\mathrm{D})}\in \mathcal{A}^{(\mathrm{V})}_{\mathrm{D}}}\Biggl[\underset{\pi_{\mathrm{A}}}{\mathbb{E}}\biggl[-c^{(\mathrm{I})}_{i,t} + \sum_{k=t+1}^{\infty}\sum_{j \in \mathcal{V}_{\mathbf{w}}}\mathbf{U}_{j,k} \Bigl\vert\overbrace{\mathbf{s}^{(\mathrm{D})}_{\mathbf{w},t}, \mathbf{b}^{(\mathrm{D})}_{\mathbf{w},t}, \mathbf{a}_{i,t}^{(\mathrm{D})}}^{\triangleq \kappa}\biggr]\Biggr]\nonumber\\
  &\numeq{c}\argmax_{\mathbf{a}_{i,t}^{(\mathrm{D})}\in \mathcal{A}^{(\mathrm{V})}_{\mathrm{D}}}\Biggl[\underset{\pi_{\mathrm{A}}}{\mathbb{E}}\biggl[-c^{(\mathrm{I})}_{i,t} + \sum_{k=t+1}^{\tau}\sum_{j \in \mathcal{V}_{\mathbf{w}}}\mathbf{U}_{j,k}\Bigl\vert\kappa\biggr]\Biggr]\nonumber\\
&\numeq{d}\argmax_{\mathbf{a}_{i,t}^{(\mathrm{D})}\in \mathcal{A}^{(\mathrm{V})}_{\mathrm{D}}}\Biggl[\underset{\pi_{\mathrm{A}}}{\mathbb{E}}\biggl[-c^{(\mathrm{I})}_{i,t} + \sum_{k=t+1}^{\tau}\sum_{j \in \mathrm{an}(i)}\mathbf{U}_{j,k}\Bigl\vert\kappa\biggr]\Biggr]\nonumber\\
&\numeq{e} \argmax_{\mathbf{a}_{i,t}^{(\mathrm{D})}\in \mathcal{A}^{(\mathrm{V})}_{\mathrm{D}}}\Biggl[\underset{\pi_{\mathrm{A}}}{\mathbb{E}}\biggl[-c^{(\mathrm{I})}_{i,t} + \sum_{k=t+1}^{\tau}\sum_{j \in \mathrm{an}(i)} u^{(\mathrm{W})}_{j,k} - c_{j,k}^{(\mathrm{I})}\Bigl\vert\kappa\biggr]\Biggr]\nonumber\\
  &\numeq{f} \argmax_{\mathbf{a}_{i,t}^{(\mathrm{D})}\in \mathcal{A}^{(\mathrm{V})}_{\mathrm{D}}}\Biggl[\underset{\pi_{\mathrm{A}}}{\mathbb{E}}\biggl[\overbrace{-c_{i,t}^{(\mathrm{I})} + \sum_{k=t+1}^{\tau}|\mathrm{an}(i)|\alpha_{i,t+1}-c_{i,k}^{(\mathrm{I})}}^{\triangleq \omega}\Bigl\vert\kappa\biggr]\Biggr] \nonumber\\
&\numeq{g} \argmax_{\mathbf{a}_{i,t}^{(\mathrm{D})}\in \mathcal{A}^{(\mathrm{V})}_{\mathrm{D}}}\Biggl[\underset{\pi_{\mathrm{A}}}{\mathbb{E}}\biggl[\omega \text{ }\Bigl\vert \text{ }\mathbf{s}_{i,t}^{(\mathrm{D})}, \mathbf{b}_{i,t}^{(\mathrm{D})}, \mathbf{a}_{i,t}^{(\mathrm{D})}\biggr]\Biggr]
    \label{eq:lemma_3_proof}
\end{align}
where $\mathbf{U}_t$ denotes the vector of utilities for all nodes at time $t$. (a) holds because $(\mathbf{U}_{j,t})_{j \in \mathcal{V}\setminus \{i\}}$ and $u^{(\mathrm{W})}_{i,t}$ are independent of $\mathbf{a}_{i,t}^{(\mathrm{D})}$ and therefore does not affect the maximization; (b) follows from the utility independence across workflows (Thm. \ref{thm:decomp}.A) and the definition of the value function $\mathscr{V}$ \cite[Thm. 7.4.1]{krishnamurthy_2016}; (c) holds because any $\mathbf{a}^{(\mathrm{D})}_{i,t}$ except $\bot$ leads to $\mathbf{s}^{(\mathrm{A})}_{i,t+1}=(0,0)$, which means that all state variables at time $k>\tau$ are independent of $\mathbf{a}_{i,t}^{(\mathrm{D})}$ and can therefore be moved outside the $\argmax$ operator; (d) follows because $(\mathbf{U}_{j,t})_{j \in \mathcal{V} \setminus \mathrm{an}(i)}$ is independent of $\mathbf{a}_{i,t}^{(\mathrm{D})}$; (e) is an expansion of $(\mathbf{U}_{j,k})_{j\in \mathrm{an}(i),k\in \{t+1,\hdots,\tau\}}$ based on (\ref{eq:objective_fun}); and (f)-(g) follow because the terms in $(u^{(\mathrm{W})}_{j,k})_{j \in \mathrm{an}(i),k \in \{t+1,\hdots,\tau\}}$ that depend on $\mathbf{a}^{(\mathrm{D})}_{i,t}$ equal $k|\mathrm{an}(i)|\alpha_{t+1,i}$, where $k$ is the constant of proportionality (see \S \ref{sec:system_model}). (Recall that $\alpha_{i,t}=1$ if node $i$ is active at time $t$ and $\alpha_{i,t}=0$ otherwise.)

The final expression in (\ref{eq:lemma_3_proof}) depends only on local information related to node $i$. This means that we can use it to define utility functions of the subgames $(\Gamma^{(i)})_{i \in \mathcal{V}_{\mathbf{w}}}$ such that they become utility-independent. Further, since the maximizer of the final expression in (\ref{eq:lemma_3_proof}) is also a maximizer of the first expression, it follows that a a best response in $\Gamma^{(i)}$ is also a best response for node $i$ in $\Gamma^{(\mathbf{w})}$ and thus in $\Gamma$ (Thm. \ref{thm:decomp}.A). Hence the subgames $(\Gamma^{(i)})_{i \in \mathcal{V}_{\mathbf{w}}}$ have optimal substructure. \qed
\subsection{Proof of Theorem \ref{thm:decomp}.C}\label{sef:proof_thm_2_c}
The idea behind this proof is that the problem of selecting \textit{which} defensive action to apply in a subgame $\Gamma^{(i)}$ (Thm. \ref{thm:decomp}.B) against a given attacker strategy can be separated from the problem of deciding \textit{when} to apply it. Through this separation, we can analyze the latter problem using optimal stopping theory. Applying a recent result by Krishnamurthy \cite[Thm. 12.3.4]{krishnamurthy_2016}, the optimal stopping strategy in $\Gamma^{(i)}$ can be characterized by switching curves.

We perform the above separation by decomposing $\mathbf{a}^{(\mathrm{D})}_{i,t}$ into two subactions: $\mathbf{a}^{(\mathrm{D},1)}_{i,t}$ and $\mathbf{a}^{(\mathrm{D},2)}_{i,t}$ which realize $\mathbf{A}^{(\mathrm{D},1)}_{i,t}$ and $\mathbf{A}^{(\mathrm{D},2)}_{i,t}$. The first subaction $\mathbf{a}^{(\mathrm{D},1)}_{i,t}\neq \bot$ determines the defensive action and the second subaction $\mathbf{a}^{(\mathrm{D},2)}_{i,t} \in \{\mathrm{S},\mathrm{C}\}$ determines when to take it. Specifically, if $\mathbf{a}^{(\mathrm{D},2)}_{i,t}=\mathrm{C}$, then $\mathbf{a}^{(\mathrm{D})}_{i,t}=\bot$, otherwise $\mathbf{a}^{(\mathrm{D})}_{i,t}=\mathbf{a}^{(\mathrm{D},1)}_{i,t}$. Using this action decomposition, at each time $t$, a strategy $\pi_{\mathrm{D}}^{(i)}$ in $\Gamma^{(i)}$ is a joint distribution over $\mathbf{A}^{(\mathrm{D},1)}_{i,t}$ and $\mathbf{A}^{(\mathrm{D},2)}_{i,t}$, which means that it can be represented in an auto-regressive manner as
\begin{align}
&\pi_{\mathrm{D}}^{(i)}(\mathbf{a}^{(\mathrm{D},1)}_{i,t}, \mathbf{a}^{(\mathrm{D},2)}_{i,t} \mid \mathbf{h}^{(\mathrm{k})}_{i,t}) \label{eq:stopping_1}\\
  &\numeq{a} \pi_{\mathrm{D}}^{(i)}(\mathbf{a}^{(\mathrm{D},1)}_{i,t} \mid \mathbf{h}^{(\mathrm{D})}_{i,t})\pi_{\mathrm{D}}^{(i)}(\mathbf{a}^{(\mathrm{D},2)}_{i,t} \mid  \mathbf{h}^{(\mathrm{D})}_{i,t}, \mathbf{a}^{(\mathrm{D},1)}_{i,t})\nonumber\\
&\numeq{b} \pi_{\mathrm{D}}^{(i)}(\mathbf{a}^{(\mathrm{D},1)}_{i,t} \mid \mathbf{b}^{(\mathrm{D})}_{i,t},\mathbf{s}^{(\mathrm{D})}_{i,t})\pi_{\mathrm{D}}^{(i)}(\mathbf{a}^{(\mathrm{D},2)}_{i,t} \mid \mathbf{b}^{(\mathrm{D})}_{i,t}, \mathbf{s}^{(\mathrm{D})}_{i,t}, \mathbf{a}^{(\mathrm{D},1)}_{i,t})  \nonumber \\
&\numeq{c} \pi_{\mathrm{D}}^{(i)}(\mathbf{a}^{(\mathrm{D},1)}_{i,t} \mid \mathbf{s}^{(\mathrm{D})}_{i,t})\pi_{\mathrm{D}}^{(i)}(\mathbf{a}^{(\mathrm{D},2)}_{i,t} \mid \mathbf{b}^{(\mathrm{D})}_{i,t}, \mathbf{s}^{(\mathrm{D})}_{i,t}, \mathbf{a}^{(\mathrm{D},1)}_{i,t})\nonumber
\end{align}
where (a) follows from the chain rule of probability; (b) holds because $(\mathbf{s}^{(\mathrm{D})}_{i,t},\mathbf{b}^{(\mathrm{D})}_{i,t})$ is a sufficient statistic for $\mathbf{H}^{(\mathrm{D})}_{i,t}$ \cite[Thm 7.2.1]{krishnamurthy_2016}; and (c) follows because
\begin{align}
&\argmax_{\mathbf{a}^{(\mathrm{D},1)}_{i,t} \in \mathcal{A}^{(\mathrm{V})}_{\mathrm{D}}\setminus \bot}\Biggl[\eta |\mathrm{an}(i)|\alpha_{i,t+1} - c^{(\mathrm{I})}_{i,t}(v^{(\mathrm{I})}_{i,t}, \mathbf{a}^{(\mathrm{D},1)}_{i,t}) + \label{eq:belief_indep}\\
  &\quad\quad\quad\quad\quad\gamma \mathbb{E}\biggl[\mathscr{V}(\mathbf{S}^{(\mathrm{D})}_{i,t+1}, \mathbf{B}^{(\mathrm{D})}_{i,t+1})\text{ }\Bigl\vert\text{ } \mathbf{s}^{(\mathrm{D})}_{i,t},\mathbf{b}^{(\mathrm{D})}_{i,t}, \mathbf{a}^{(\mathrm{D},1)}_{i,t}\biggr]\Biggr]\nonumber\\
&\numeq{a}\argmax_{\mathbf{a}^{(\mathrm{D},1)}_{i,t} \in \mathcal{A}^{(\mathrm{V})}_{\mathrm{D}}\setminus \bot}\Biggl[\eta |\mathrm{an}(i)|\alpha_{i,t+1} - c^{(\mathrm{A})}(\mathbf{a}^{(\mathrm{D},1)}_{i,t}) + \nonumber\\
  &\quad\quad\quad\quad\quad\gamma \mathbb{E}\biggl[\mathscr{V}(\mathbf{S}^{(\mathrm{D})}_{i,t+1}, \mathbf{B}^{(\mathrm{D})}_{i,t+1})\text{ }\Bigl\vert\text{ } \mathbf{s}^{(\mathrm{D})}_{i,t},\mathbf{b}^{(\mathrm{D})}_{i,t}, \mathbf{a}^{(\mathrm{D},1)}_{i,t}\biggr]\Biggr]\nonumber\\
  &\numeq{b}\argmax_{\mathbf{a}^{(\mathrm{D},1)}_{i,t} \in \mathcal{A}^{(\mathrm{V})}_{\mathrm{D}}\setminus \bot}\Biggl[\eta |\mathrm{an}(i)|\alpha_{i,t+1} - c^{(\mathrm{A})}(\mathbf{a}^{(\mathrm{D},1)}_{i,t}) \nonumber\\
&\quad\quad\quad\quad\quad\quad\quad\gamma \mathbb{E}\biggl[\mathscr{V}(\mathbf{S}^{(\mathrm{D})}_{i,t+1}, \mathbf{e}_1)\text{ }\Bigl\vert\text{ } \mathbf{s}^{(\mathrm{D})}_{i,t}, \mathbf{a}^{(\mathrm{D},1)}_{i,t}\biggr]\Biggr]\nonumber
\end{align}
which means that $\mathbf{a}^{(\mathrm{D},1)}_{i,t}\neq \bot$ is independent of $\mathbf{B}_{i,t}^{(\mathrm{D})}$. The first statement in (\ref{eq:belief_indep}) is the Bellman equation \cite[Eq. 1]{bellman1957markovian}; (a) holds because $V^{(\mathrm{I})}_{i,t}$ is independent of $\mathbf{a}^{(\mathrm{D},1)}_{i,t}$; and (b) is true because any $\mathbf{a}^{(\mathrm{D},1)}_{i,t}\neq \bot$ leads to $\mathbf{S}^{(\mathrm{A})}_{i,t+1}=(0,0)$ and thus to $\mathbf{B}^{(\mathrm{D})}_{i,t+1}=\mathbf{e}_1=(1,0,0)$. (Recall that the belief space $\mathcal{B}^{(i)}_{\mathrm{D}}$ is the two-dimensional unit simplex.)

The strategy decomposition in (\ref{eq:stopping_1}) means that we can obtain a best response strategy in $\Gamma^{(i)}$ by jointly optimizing two substrategies: $\pi_{\mathrm{D}}^{(i,1)}$ and $\pi_{\mathrm{D}}^{(i,2)}$. The former corresponds to solving an \mdp{} $\mathscr{M}^{(\mathrm{D},1)}$ with state space $\mathbf{s}_{i}^{(\mathrm{D})} \in \mathcal{Z}$ and the latter corresponds to solving a set of optimal stopping \pomdp{}s $(\mathscr{M}^{(\mathrm{D},2)}_{i,s^{(\mathrm{D})},a^{(\mathrm{D})}})_{s^{(\mathrm{D})} \in \mathcal{Z}, a^{(\mathrm{D})} \in \mathcal{A}_{\mathrm{D}}^{(\mathrm{V})}}$ with state space $\mathbf{s}_{i}^{(\mathrm{A})} \in \{(0,0),(1,0),(1,1)\}=\{0,1,2\}$.

Each stopping problem can be defined with a \textit{single} stop action rather than multiple stop actions \cite[\S III.C]{hammar_stadler_tnsm} because
\begin{align}
&\argmax_{\pi_{\mathrm{D}} \in \Pi^{(i,2)}_{\mathrm{D}}}\Biggl[\mathbb{E}_{\pi_{\mathrm{D}}}\biggl[\sum_{t=1}^{\infty}\gamma^{t-1}\mathbf{U}_{i,2,t} \text{ }\Bigl\vert\text{ } \mathbf{B}^{(\mathrm{D})}_{i,1}=\mathbf{e}_1\biggr]\Biggr] \nonumber\\
              &=\argmax_{\pi_{\mathrm{D}} \in \Pi^{(i,2)}_{\mathrm{D}}}\Biggl[\mathbb{E}_{\pi_{\mathrm{D}}}\biggl[\sum_{t=1}^{\tau_1}\gamma^{t-1}\mathbf{U}_{i,2,t} \text{ }\Bigl\vert\text{ } \mathbf{B}^{(\mathrm{D})}_{i,1}=\mathbf{e}_1\biggr] +\nonumber\\
  &\quad\quad\quad\mathbb{E}_{\pi_{\mathrm{D}}}\biggl[\sum_{t=\tau_1+1}^{\tau_2}\gamma^{t-1}\mathbf{U}_{i,2,t} \text{ }\Bigl\vert\text{ } \mathbf{B}^{(\mathrm{D})}_{i,\tau_1+1}=\mathbf{e}_1\biggr] + \hdots \Biggr]\nonumber\\
&=\argmax_{\pi_{\mathrm{D}} \in \Pi^{(i,2)}_{\mathrm{D}}}\Biggl[\mathbb{E}_{\pi_{\mathrm{D}}}\biggl[\sum_{t=1}^{\tau_1}\gamma^{t-1}\mathbf{U}_{i,2,t} \text{ }\Bigl\vert\text{ } \mathbf{B}^{(\mathrm{D})}_{i,1}=\mathbf{e}_1\biggr]\Biggr] \label{eq:stopping_indep}
\end{align}
where $\Pi^{(i,2)}_{\mathrm{D}}$, $\mathbf{U}_{i,2,t}$, and $\tau_1, \tau_2, \hdots $ denote the strategy space, utility, and stopping times in $\mathscr{M}^{(\mathrm{D},2)}_{i,s^{(\mathrm{D})},a^{(\mathrm{D})}}$. Note that the belief space $\mathcal{B}^{(i)}_{\mathrm{D}}$ for each stopping problem is the $2$-dimensional unit simplex and that $\mathbf{B}^{(\mathrm{D})}_{i,\tau_j+1}=\mathbf{e}_1=(1,0,0)$ for each stopping time $\tau_j$ since $\mathbf{a}^{(\mathrm{D},2)}_{i,\tau_j}=\mathrm{S}\implies \mathbf{s}^{(\mathrm{A})}_{i,\tau_j+1}=(0,0)$.

The transition matrices for each stopping problem are of the form:
\begin{align}
\begin{bmatrix}
1-p & p & 0\\
0 & 1-q & q\\
0 & 0 & 1
\end{bmatrix}
\quad \text{and}\quad
\begin{bmatrix}
1 & 0 & 0\\
1 & 0 & 0\\
1 & 0 & 0
\end{bmatrix}\label{eq:p_tp2}
\end{align}
where $p$ is the probability that the attacker performs reconnaissance and $q$ is the probability that the attacker compromises the node. The left matrix in (\ref{eq:p_tp2}) relates to $\mathbf{a}_{i,t}^{(\mathrm{D},2)}=\mathrm{C}$ and the right matrix relates to $\mathbf{a}_{i,t}^{(\mathrm{D},2)}=\mathrm{S}$. The non-zero second order minors of the matrices are $(1-p)(1-q)$, $pq$, $1-q$, $1-p$, $p$, and $(1-p)q$, which implies that the matrices are \tpp \cite[Def. 10.2.1]{krishnamurthy_2016}. Further, it follows from (\ref{eq:objective_fun}) that the utility function of the stopping problems satisfies
\begin{align*}
\mathbf{u}_{j,2}(0,a) \geq \mathbf{u}_{j,2}(1,a) \geq \mathbf{u}_{j,2}(2,a)
\end{align*}
for $a \in \{\mathrm{S}, \mathrm{C}\}$ and that
\begin{align*}
\mathbf{u}_{j,2}(s,\mathrm{S})-\mathbf{u}_j(s,\mathrm{C}) &\geq \mathbf{u}_{j,2}(2,\mathrm{S})-\mathbf{u}_{j,2}(2,\mathrm{C})\\
\mathbf{u}_{j,2}(0,\mathrm{S})-\mathbf{u}_j(0,C) &\geq \mathbf{u}_{j,2}(s,\mathrm{S})-\mathbf{u}_{j,2}(s,\mathrm{C})
\end{align*}
for all $s \in \mathcal{S}^{(j),2}$. Since the distributions $Z_{\mathbf{O}_1\mid \mathbf{s}^{(\mathrm{A})}},\hdots, Z_{\mathbf{O}_{|\mathcal{V}|}\mid \mathbf{s}^{(\mathrm{A})}}$ are \tpp by assumption it then follows from \cite[Thm. 12.3.4]{krishnamurthy_2016} that there exists a switching curve $\Upsilon$ that partitions $\mathcal{B}_{\mathrm{D}}^{(i)}$ into two individually connected regions: a stopping set $\mathscr{S}^{(i)}_{\mathrm{D}}$ where $\mathbf{a}_{i,t}^{(\mathrm{D},2)}=\mathrm{S}$ is a best response and a continuation set $\mathscr{C}^{(i)}_{\mathrm{D}}$ where $\mathbf{a}_{i,t}^{(\mathrm{D},2)}=\mathrm{C}$ is a best response (see Fig. \ref{fig:unit_simplex_2}).

The argument behind the existence of a switching curve is as follows \cite[Thm. 12.3.4]{krishnamurthy_2016}. On any line segment $\mathcal{L}(\mathbf{e}_1,\widehat{\mathbf{b}}^{(\mathrm{D})})$ in $\mathcal{B}_{\mathrm{D}}^{(i)}$ that starts at $\mathbf{e}_1$ and ends at the subsimplex joining $\mathbf{e}_2$ and $\mathbf{e}_3$ (denoted with $\widehat{\mathbf{b}}^{(\mathrm{D})} \in \mathcal{B}_{\mathrm{D},\mathbf{e}_1}^{(i)}$), all belief states are totally ordered with respect to the Monotone Likelihood Ratio (\mlr) order \cite[Def. 10.1.1]{krishnamurthy_2016}. As a consequence, Topkis's theorem \cite[Thm. 6.3]{topkis_theorem} implies that the optimal strategy on $\mathcal{L}(\mathbf{e}_1,\widehat{\mathbf{b}}^{(\mathrm{D})})$ is monotone with respect to the \mlr order. Consequently, there exists a threshold belief state $\alpha_{\widehat{\mathbf{b}}^{(\mathrm{D})}}$ on $\mathcal{L}(\mathbf{e}_1,\widehat{\mathbf{b}}^{(\mathrm{D})})$ where the optimal strategy switches from $\mathrm{C}$ to $\mathrm{S}$. Since $\mathcal{B}_{\mathrm{D}}^{(i)}$ can be covered by the union of lines $\mathcal{L}(\mathbf{e}_1,\widehat{\mathbf{b}}^{(\mathrm{D})})$, the thresholds $\alpha_{\widehat{\mathbf{b}}^{(\mathrm{D})}_1},\alpha_{\widehat{\mathbf{b}}^{(\mathrm{D})}_2},\hdots$ yield a switching curve $\Upsilon$. \qed
\section{Finding Nash Equilibria of the\\ Decomposed Intrusion Response Game}\label{sec:rl_method}
To find a Nash equilibrium of $\Gamma$ (\ref{def:game}) we develop a \textit{fictitious self-play} algorithm called Decompositional Fictitious Self-Play (\dfsp), which estimates Nash equilibria based on the decomposition presented above. The pseudocode is listed in Alg. \ref{alg:decomposition}. (In Alg. \ref{alg:decomposition}, $\oplus$ denotes vector concatenation, $-\mathrm{k}$ denotes the opponent of player $\mathrm{k}$, and $\mathscr{M}^{(\mathrm{k})}_{i}$ denotes the best response \pomdp of $\mathrm{k}$ in $\Gamma^{(i)}$ (Thm \ref{thm:decomp}).)

\begin{algorithm}
  \SetNoFillComment
  \SetKwProg{myInput}{Input:}{}{}
  \SetKwProg{myOutput}{Output:}{}{}
  \SetKwFunction{pomdpsolver}{\textsc{pomdpsolver}}
  \SetKwProg{myalg}{Algorithm}{}{}
  \SetKwProg{myproc}{Procedure}{}{}
  \SetKw{KwTo}{inp}
  \SetKwFor{Forp}{in parallel for}{\string do}{}%
  \SetKwProg{mylamb}{\textbf{$\lambda$}}{}{}
  \SetKwFor{Loop}{Loop}{}{EndLoop}
  \DontPrintSemicolon
  \SetKwBlock{DoParallel}{do in parallel}{end}
  \myInput{
    \textsc{p-solver}: \upshape a \pomdp solver,\\
    \quad \quad\quad \text{ }$\delta$:\upshape \text{ }convergence criterion, $\Gamma$: the \poposg
  }{}
  \myOutput{
    \upshape An approximate Nash equilibrium $(\pi_{\mathrm{D}}, \pi_{\mathrm{A}})$
  }{}
  \caption{\dfsp}\label{alg:decomposition}
  \myalg{\dfsp(\textsc{p-solver}, $\delta$, $\Gamma$)}{
    Initialize $\pi_{\mathrm{D}},\pi_{\mathrm{A}},\widehat{\delta}$\;
    \While{$\widehat{\delta} \geq \delta$}{
      \Forp{$\mathrm{k} \in \{\mathrm{D}, \mathrm{A}\}$}{
        $\bm{\pi}_{\mathrm{k}} \leftarrow $\textsc{local-brs}$(\textsc{p-solver}, \Gamma, \mathrm{k}, \pi_{-\mathrm{k}})$\;
        $\tilde{\pi}_{\mathrm{k}} \leftarrow $\textsc{composite-strategy}$(\Gamma, \bm{\pi}_{\mathrm{k}})$\;
        $\pi_{\mathrm{k}} \leftarrow $\textsc{average-strategy}$(\pi_{\mathrm{k}}, \tilde{\pi}_{\mathrm{k}})$\;
      }
      $\widehat{\delta} \leftarrow $\textsc{exploitability}($\tilde{\pi}_{\mathrm{D}}$,$\tilde{\pi}_{\mathrm{A}}$) (see (\ref{eq:approx_exp}))
    }
  \Return $(\pi_{\mathrm{D}}, \pi_{\mathrm{A}})$
  }
  \myproc{\textsc{local-brs}$(\textsc{p-solver}, \Gamma, \mathrm{k}, \pi_{-\mathrm{k}})$}{
    $\bm{\pi}_{\mathrm{k}} \leftarrow ()$\;
    \Forp{$\mathbf{w} \in \mathcal{W}$, $(i) \in \mathcal{V}_{\mathbf{w}}$}{
        $\bm{\pi}_{\mathrm{k}} \leftarrow \bm{\pi}_{\mathrm{k}} \oplus $ \textsc{p-solver}($\mathscr{M}^{(\mathrm{k})}_{i}, \pi_{-\mathrm{k}}$)
    }
    \Return $\bm{\pi}_{\mathrm{k}}$
  }
  \myproc{\textsc{composite-strategy}$(\Gamma, \bm{\pi}_{\mathrm{k}})$}{
    \Return $\pi_{\mathrm{k}} \leftarrow $\text{\textbf{Procedure }}\mylamb{($\mathbf{s}_t^{(\mathrm{k})}$, $\mathbf{b}_t^{(\mathrm{k})}$)}{
      $\mathbf{a}^{(\mathrm{k})}_t \leftarrow ()$\;
      \For{$\mathbf{w} \in \mathcal{W}$, $i \in \mathcal{V}_{\mathbf{w}}$}{
          $\mathbf{a}^{(\mathrm{k})}_t \leftarrow \mathbf{a}^{(\mathrm{k})}_t \oplus (\bm{\pi}_{\mathrm{k}}^{(i)}(\mathbf{s}_{i,t}^{(\mathrm{k})}, \mathbf{b}_{i,t}^{(\mathrm{k})}))$\;
      }
      \Return $\mathbf{a}^{(\mathrm{k})}_t$
    }
  }
\end{algorithm}

\dfsp implements the fictitious play process described in \cite{brown_fictious_play} and generates a sequence of strategy profiles $(\pi_{\mathrm{D}}, \pi_{\mathrm{A}})$, $(\pi^{\prime}_{\mathrm{D}}$, $\pi^{\prime}_{\mathrm{A}})$, $\hdots$ that converges to a Nash equilibrium $(\pi^{*}_{\mathrm{D}}, \pi^{*}_{\mathrm{A}})$ \cite[Thms. 7.2.4--7.2.5]{multiagent_systems_book_1}. During each step of this process, \dfsp learns best responses against the players' current strategies and then updates both players' strategies (lines $7$--$11$ in Alg. \ref{alg:decomposition}). To obtain the best responses, it first finds best responses for the node subgames as constructed in the proof of Thm. \ref{thm:decomp}.B (lines $14$--$18$), and then it combines them using the method described in \S \ref{proof:thm_2_b} (lines $19$--$25$).

Finding best responses for node subgames amounts to solving \pomdps. The principal method for solving \pomdps is dynamic programming \cite{krishnamurthy_2016}. Dynamic programming is however intractable in our case, as demonstrated in Fig. \ref{fig:pomdp_scale}. To find the best responses we instead resort to approximation algorithms. More specifically, we use the Proximal Policy Optimization (\ppo) algorithm \cite[Alg. 1]{ppo} to find best responses for the attacker, and we use a combination of dynamic programming and stochastic approximation to find best responses for the defender. In particular, to find best responses for the defender, we first solve the \mdp{} defined in \S \ref{sef:proof_thm_2_c} via the value iteration algorithm \cite[Eq. 6.21]{krishnamurthy_2016}, which can be done efficiently due to full observability. After solving the \mdp{}, we approximate the optimal switching curves defined in the proof of Thm. \ref{thm:decomp}.C (\S \ref{sef:proof_thm_2_c}) with the following linear approximation \cite[Eq. 12.18]{krishnamurthy_2016}.
\begin{align}
\pi_{\mathrm{D}}(\mathbf{b}^{(\mathrm{D})}) &=
\begin{dcases}
  \mathrm{S} & \text{if }
  \begin{bmatrix}
    0 & 1 & \bm{\theta}
  \end{bmatrix}
  \begin{bmatrix}
    (\mathbf{b}^{(\mathrm{D})})^T\\
    -1
  \end{bmatrix}
  > 0
  \\
  \mathrm{C} & \text{otherwise}
\end{dcases}\label{eq:linear_approx} \\
&\text{subject to } \bm{\theta} \in \mathbb{R}^2,\text{ }\bm{\theta}_2 > 0,\text{ and }\bm{\theta}_1 \geq 1 \nonumber
\end{align}
The coefficients $\bm{\theta}$ in (\ref{eq:linear_approx}) are estimated through the stochastic approximation algorithm in \cite[Alg. 14]{krishnamurthy_2016} and \cite[Alg. 1]{hammar_stadler_tnsm}.

\begin{figure}
  \centering
  \scalebox{0.95}{
    \input{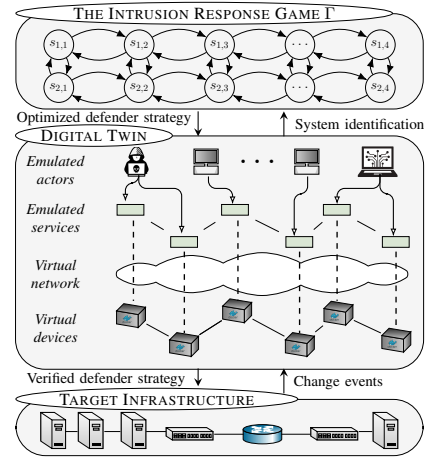}
  }
  \caption{The digital twin is a virtual replica of the target infrastructure and is used in our framework for evaluation and data collection \cite{csle_docs,digital_twins_kim}.}
  \label{fig:digital_twin}
\end{figure}
\begin{figure*}
  \centering
    \scalebox{0.373}{
      \includegraphics{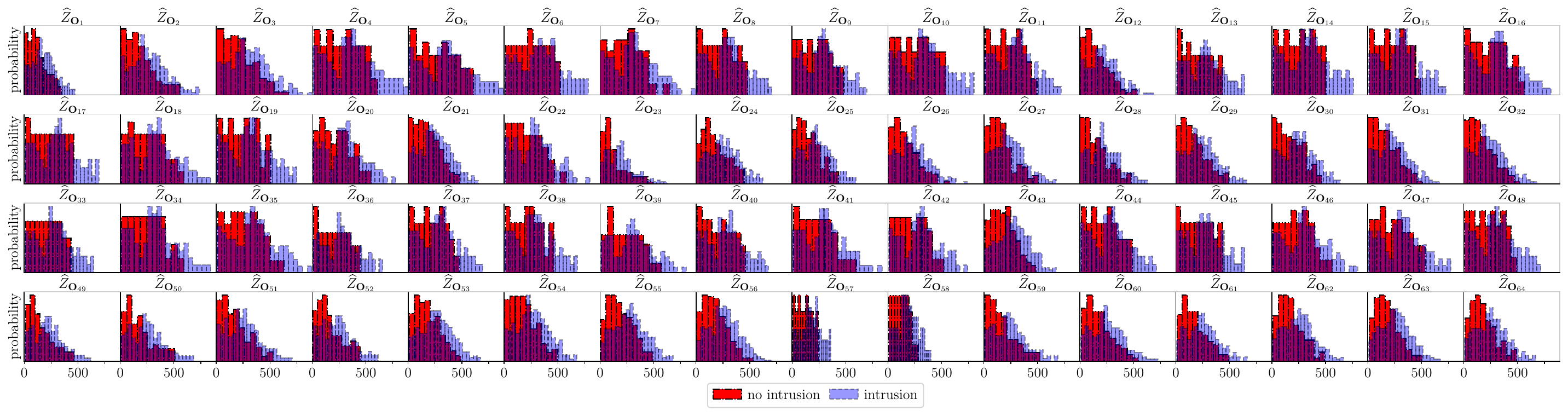}
    }
    \caption{Empirical observation distributions $\widehat{Z}_{\mathbf{O}_1}, \hdots, \widehat{Z}_{\mathbf{O}_{|\mathcal{V}|}}$ as estimates of $Z_{\mathbf{O}_1}, \hdots, Z_{\mathbf{O}_{|\mathcal{V}|}}$ in the target infrastructure (depicted in Fig. \ref{fig:use_case}); $\mathbf{O}_{i}$ is a random variable representing the number of \idps alerts related to node $i\in \mathcal{V}$, weighted by priority; the x-axes show the local observation spaces $\mathcal{O}^{(\mathrm{V})}$ for each node; the y-axes show $Z(\mathbf{O}_{i} \mid \mathbf{S}_{i})$ (\ref{eq:obs_1}).}
    \label{fig:observation_distributions}
  \end{figure*}
\section{Digital Twin and System Identification}\label{sec:dt_and_identification}
The \dfsp algorithm described above approximates a Nash equilibrium of $\Gamma$ (\ref{def:game}) by simulating games and updating both players' strategies through reinforcement learning and dynamic programming. To identify the parameters required to instantiate these simulations and to evaluate the learned strategies, we use a digital twin of the target infrastructure (see Fig. \ref{fig:digital_twin}). This section describes the digital twin (\S \ref{sec:dt}) and the identification process (\S \ref{sec:identification}).
\subsection{Creating a Digital Twin of the Target Infrastructure}\label{sec:dt}
We create a digital twin of the target infrastructure shown in Fig. \ref{fig:use_case} through an emulation system. Documentation of this emulation system is available in \cite{csle_docs,digital_twins_kim}; the source code is available at \cite{csle_source_code}; and a video demonstration is available at \cite{video_demonstration3}.

The process of creating the digital twin involves two main tasks. The first task is to replicate relevant parts of the physical infrastructure that is emulated, such as physical resources, network interfaces, and network conditions. This task is described in \S \ref{sec:emulation_method}. The second task is to emulate actors in the digital twin (e.g., the attacker, the defender, and the client population). We describe this task in \S \ref{sec:emulating_actors}.
\subsubsection{Emulating physical resources}\label{sec:emulation_method}
The physical resources of the target infrastructure are emulated through the following steps.

\vspace{1mm}

\noindent\textbf{Emulating physical hosts.} Physical hosts are emulated with Docker containers \cite{docker}, i.e., lightweight executable packages that include runtime systems, code, system tools, system libraries, and configurations. Resource allocation to containers, e.g., \cpu and memory, is enforced using \cgroups. The software functions running inside the containers replicate important components of the target infrastructure, such as, web servers, databases, the \snort \idps \cite{snort}, and the \ryu \sdn controller \cite{ryu_sdn_controller}.

\vspace{1mm}

\noindent\textbf{Emulating physical switches.} Physical switches are emulated with Docker containers that run Open vSwitch (\ovs) \cite{ovs} and may connect to a controller through the \openflow protocol \cite{openflow}. (Since the switches are programmed through flow tables, they can act either as classical layer 2 switches or as routers, depending on the flow table configurations.)

\vspace{1mm}

\noindent\textbf{Emulating physical network links.} Network connectivity is emulated with virtual links implemented by Linux bridges. Network isolation between virtual containers on the same physical host is achieved through network namespaces, which create logical copies of the physical host's network stack. To connect containers on different physical hosts, the emulated traffic is tunneled over the physical network using \vxlan tunnels \cite{vxlan}.

\vspace{1mm}

\noindent\textbf{Emulating network conditions.} Network conditions of virtual links are configured using the \netem module in the Linux kernel \cite{netem}. This module allows fine-gained configuration of bit rates, packet delays, packet loss probabilities, jitter, and packet reordering probabilities.

We emulate connections between nodes as full-duplex loss-less connections of $1$ Gbit/s capacity in both directions. We emulate connections between the gateway and the external client population as full-duplex connections of $100$ Mbit/s capacity and $0.1\%$ packet loss with random bursts of $1\%$ packet loss. (These numbers are based on measurements on enterprise and wide-area networks \cite{packet_losses_decreasing,Paxson97end-to-endinternet,elliott_markov_chain_ref}.)

\subsubsection{Emulating Actors in the Digital Twin}\label{sec:emulating_actors}
In this section, we describe how actors of the intrusion response use case described in \S \ref{sec:use_case} are emulated in the digital twin.

\vspace{1mm}

\noindent\textbf{Emulating the client population.} The \textit{client population} is emulated by processes in Docker containers. Clients interact with application nodes through the gateway by consuming workflows. The workflow of a client is selected uniformly at random and its sequence of service invocations is decided uniformly at random. Client arrivals per time-step are emulated using a stationary Poisson process with rate $\lambda=50$ and exponentially distributed service times with mean $\mu=4$. The duration of a time-step in the emulation is $30$ seconds.

\vspace{1mm}

\noindent\textbf{Emulating the attacker.} The attacker's actions are emulated by executing scripts that automate exploits (see Table \ref{tab:attacker_actions}).

\begin{table}
\centering
\begin{tabular}{ll} \toprule
  {\textit{Type}} & {\textit{Actions}} \\ \midrule
  Reconnaissance  & \tcpp \syn scan, \udp port scan, \tcpp \xmas scan\\
                  & \vulscan vulnerability scanner, ping-scan \\
  &\\
  Brute-force & \telnet, \ssh, \ftp, \cassandra,\\
                  &  \irc, \mongo, \mysql, \smtp, \postgres\\
                  &\\
  Exploit & \cve-2017-7494, \cve-2015-3306,\\
                  & \cve-2010-0426, \cve-2015-5602,\cve-2015-1427 \\
                  &  \cve-2014-6271, \cve-2016-10033\\
                  & \cwe-89 weakness on the web app DVWA \cite{dvwa}\\
  \bottomrule\\
\end{tabular}
\caption{Attacker actions executed on the digital twin; actions that exploit vulnerabilities in specific software products are identified by the vulnerability identifiers in the Common Vulnerabilities and Exposures (\cve) database \cite{cve}; actions that exploit vulnerabilities that are not described in the \cve database are categorized according to the types of the vulnerabilities they exploit based on the Common Weakness Enumeration (\cwe) list \cite{cwe}; shell commands and scripts for executing the actions are listed in \cite{supp_material_hammar_stadler}; further details about the actions can be found in Appendix \ref{appendix:attacker actions}.}\label{tab:attacker_actions}
\end{table}

\vspace{1mm}

\noindent\textbf{Emulating the defender.} The four types of defender actions (see Fig. \ref{fig:defender_actions}) are emulated as follows. To emulate the \textit{node migration} action, we remove all virtual network interfaces of the emulated node and add a new interface that connects it to the new zone. To emulate the \textit{flow migration/blocking} action we add rules to the flow tables of the emulated switches that match all flows towards the node and redirect them to a given destination. To emulate the \textit{node shut down} action, we shut down the virtual container corresponding to the emulated node. Finally, to emulate the \textit{access control} action, we reset all user accounts and certificates on the emulated node.
\subsection{Estimating the Observation Distributions}\label{sec:identification}
Following the intrusion response use case described in \S \ref{sec:use_case}, we define the observation $\mathbf{O}_{i,t}$ to be the number of \idps alerts associated with node $i$ at time $t$, weighted by priority. As our target infrastructure consists of $64$ nodes (see App. \ref{appendix:infrastructure_configuration} and Fig. \ref{fig:use_case}), there are $64$ alert distributions $Z_{\mathbf{O}_1}, \hdots, Z_{\mathbf{O}_{64}}$ (\ref{eq:obs_1}). We estimate these distributions based on empirical data from the digital twin.

\begin{figure*}
\captionsetup[subfigure]{justification=centering}
\centering
    \begin{subfigure}[t]{1\textwidth}
        \centering
    \scalebox{0.29}{
      \includegraphics{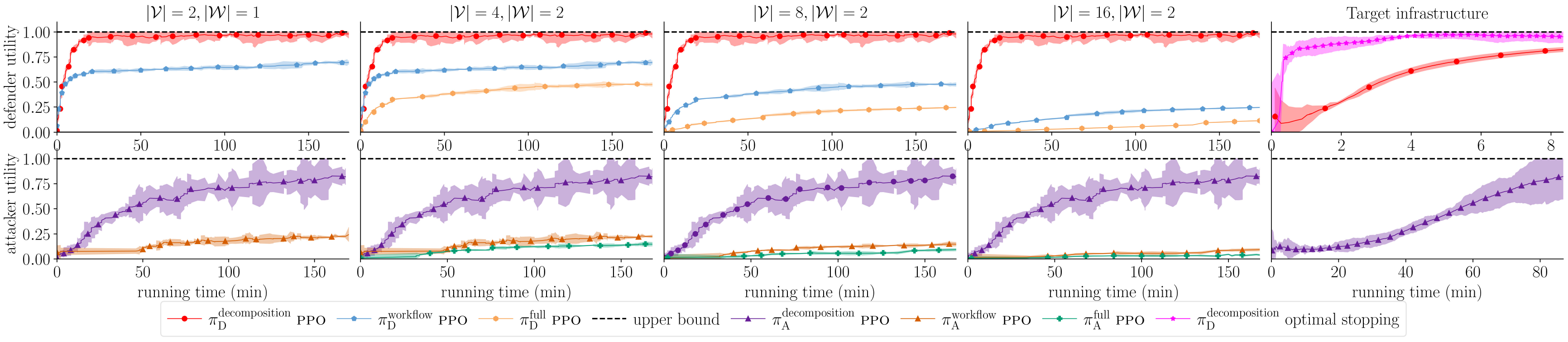}
    }
    \caption{Best response learning curves for the target infrastructure and synthetic infrastructures with varying $|\mathcal{V}|$ and $|\mathcal{W}|$.}\label{fig:best_response_ppo_returns}
  \end{subfigure}
    \begin{subfigure}[t]{0.35\textwidth}
        \centering
    \scalebox{0.7}{
      \includegraphics{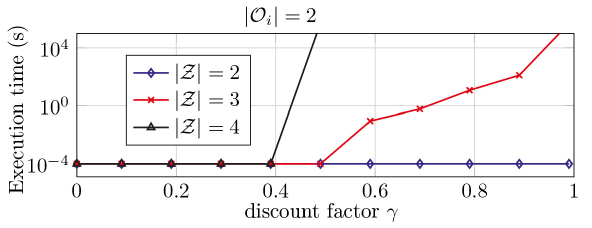}
    }
    \caption{Runtimes of dynamic programming.}\label{fig:pomdp_scale}
    \end{subfigure}
    \hfill
    \begin{subfigure}[t]{0.35\textwidth}
        \centering
    \scalebox{0.7}{
      \includegraphics{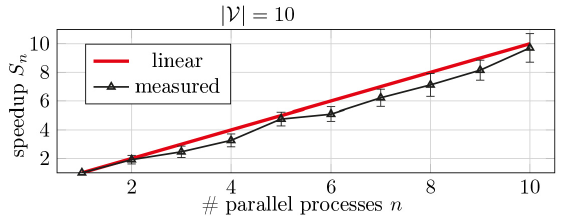}
    }
    \caption{Best response scalability.}
    \label{fig:decomposition_scale}
    \end{subfigure}
    \hfill
    \begin{subfigure}[t]{0.2\textwidth}
        \centering
  \scalebox{0.25}{
    \begin{tikzpicture}

\node[scale=0.8] (kth_cr) at (9,2.15)
{
  \begin{tikzpicture}
\draw[-, black, thick, line width=0.9mm, name path=simplex] (0,0) to (10,0) to (5,10) to (0,0);
\draw[-, black, thick, line width=0.9mm, name path=simplex2] (10,0) to (0,0) to (5,10);

\draw[line width=0.4mm, name path=switching,line width=0.9mm] (2, 4) to (3.8,0);

\node[inner sep=0pt,align=center, scale=3, rotate=0, opacity=1] (obs) at (0.75,-0.6)
{
healthy
};

\node[inner sep=0pt,align=center, scale=3, rotate=0, opacity=1] (obs) at (9.5,-0.6)
{
discovered
};
\node[inner sep=0pt,align=center, scale=3, rotate=0, opacity=1] (obs) at (5,10.45)
{
  compromised
};
\node[inner sep=0pt,align=center, scale=3, rotate=0, opacity=1] (obs) at (1.45,4.85)
{
  $0.4$
};
\node[inner sep=0pt,align=center, scale=3, rotate=0, opacity=1] (obs) at (3.8,-0.6)
{
  $0.38$
};
\path[
pattern=north west lines,
        intersection segments={
                of=switching and simplex2,
                sequence={R2--L2}
              }];

  \end{tikzpicture}
};

\end{tikzpicture}
  }
  \caption{Best response structure.}\label{fig:learned_switches}
\end{subfigure}
\caption{Best response learning via decomposition; (a) shows learning curves in simulation for synthetic infrastructures and the target infrastructure; the curves show the mean and 95\% confidence interval for five random seeds; (b) shows execution times of computing best responses via dynamic programming and Sondik's value iteration algorithm \cite{smallwood_1}; (c) shows the speedup of our approach when computing best responses with different number of parallel processes; the speedup is calculated as $S_n = \frac{T_1}{T_{n}}$ where $T_n$ is the completion time with $n$ processes; and (d) shows an estimated switching curve (Thm. \ref{thm:decomp}.C).}
    \label{fig:mega_1}
\end{figure*}
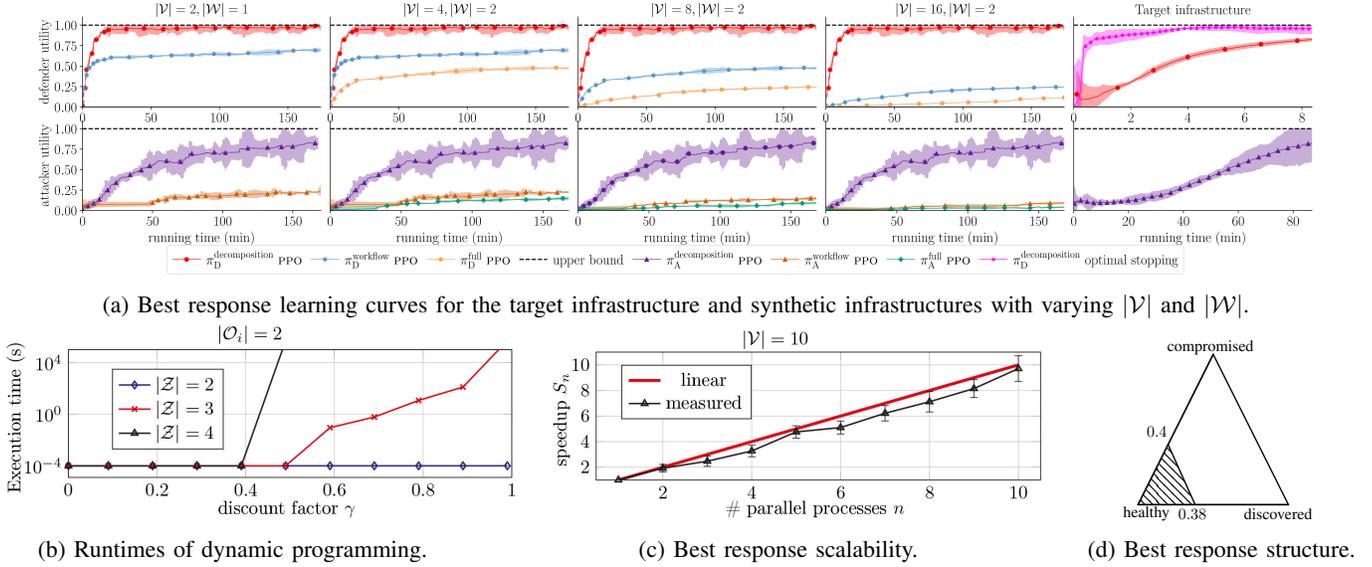

At the end of every time-step in the digital twin we collect the number of \idps alerts that occurred during the time-step. These values are then used to compute the vector $\mathbf{o}_t$, which contains the total number of \idps alerts per node, weighted by priority. For the evaluation in this paper we collect measurements from $10^4$ time-steps using the Snort \idps \cite{snort}. (Each time-step in the digital twin is $30$ seconds.) Based on these measurements, we compute the empirical distributions $\widehat{Z}_{\mathbf{O}_1}, \hdots, \widehat{Z}_{\mathbf{O}_{64}}$ as estimates of $Z_{\mathbf{O}_1}, \hdots, Z_{\mathbf{O}_{64}}$ (see Fig. \ref{fig:observation_distributions}).

We observe in Fig. \ref{fig:observation_distributions} that the distributions differ between nodes, which can be explained by the different services provided by the nodes (see App. \ref{appendix:infrastructure_configuration}). We further observe that both the distributions when no intrusion occurs and the distributions during intrusion have most of their probability masses within $[0, 300]$. The distributions during intrusion also have substantial probability mass at larger values.

\underline{Remark:} the stochastic matrices with the rows $\widehat{Z}_{\mathbf{O}_i|\mathbf{s}^{(\mathrm{A})}_i=(0,0)}$ and $\widehat{Z}_{\mathbf{O}_i|\mathbf{s}^{(\mathrm{A})}_i \neq (0,0)}$ have $250 \times 10^9$ second-order minors, which are almost all non-negative. This suggests that the \tpp assumption in Thm. \ref{thm:decomp}.C can be made.
\section{Experimental Evaluation}
Our approach to find near-optimal defender strategies includes learning Nash equilibrium strategies via the \dfsp algorithm and evaluating these strategies on the digital twin (see Fig. \ref{fig:overview}). This section describes the evaluation results.

\vspace{1mm}

\noindent\textbf{Experiment setup.} The instantiation of $\Gamma$ (\ref{def:game}) and the hyperparameters are listed in App. \ref{appendix:hyperparameters}. We evaluate \dfsp both on a digital twin of the target infrastructure and in simulations of synthetic infrastructures. The topology of the target infrastructure is depicted in Fig. \ref{fig:use_case} and its configuration is available in App. \ref{appendix:infrastructure_configuration}. The digital twin is deployed on a server with a $24$-core \textsc{intel} \textsc{xeon} \textsc{gold} \small $2.10$ GHz \normalsize \textsc{cpu} and $768$ \textsc{gb} \textsc{ram}. Simulations of $\Gamma$ and executions of \dfsp run on a cluster with $2$x\textsc{tesla} \textsc{p}\small 100 \normalsize\textsc{gpus}, $4$x\textsc{rtx}\small$8000$ \normalsize \textsc{gpus}, and $3$x\small$16$-core \textsc{intel} \textsc{xeon} $3.50$ GHz \normalsize \textsc{cpus}. Code for replicating the experiments is available in \cite{csle_docs,csle_source_code}.

\vspace{1mm}

\noindent\textbf{Convergence metric.} To estimate the convergence of the sequence of strategy pairs generated by \dfsp, we use the \textit{approximate exploitability} metric $\widehat{\delta}$ \cite{approx_br}:
\begin{align}
\widehat{\delta} = \mathbb{E}_{\widehat{\pi}_{\mathrm{D}},\pi_{\mathrm{A}}}\left[J\right] - \mathbb{E}_{\pi_{\mathrm{D}},\widehat{\pi}_{\mathrm{A}}}\left[J\right] \label{eq:approx_exp}
\end{align}
where $J$ is defined in (\ref{eq:objective_fun}) and $\widehat{\pi}_{\mathrm{k}}$ denotes an approximate best response strategy for player $\mathrm{k}$. The closer $\widehat{\delta}$ becomes to $0$, the closer $(\pi_{\mathrm{D}},\pi_{\mathrm{A}})$ is to a Nash equilibrium.

\vspace{1mm}

\noindent\textbf{Baseline algorithms.} We compare the performance of our approach ($\pi^{\text{decomposition}}$) with two baselines: $\pi^{\text{full}}$ and $\pi^{\text{workflow}}$. Baseline $\pi^{\text{full}}$ solves the full game without decomposition and $\pi^{\text{workflow}}$ decomposes the game on the workflow-level only.

We compare the performance of \dfsp with that of Neural Fictitious Self-Play (\nfsp) \cite[Alg. 1]{heinrich_1} and \ppo \cite[Alg.1 ]{ppo}, which are the most popular algorithms among related work (see \cite[\S VII]{hammar_stadler_game_23} for a review of algorithms used in related work).

\vspace{1mm}

\noindent\textbf{Baseline strategies.} We compare the defender strategies learned through \dfsp with three baselines. The first baseline selects actions uniformly at random. The second baseline assumes prior knowledge of the opponent's actions and acts optimally based on this information. The last baseline acts according to the following heuristic: shut down a node $i \in \mathcal{V}$ when an \idps alert occurs, i.e., when $\mathbf{o}_{i,t} > 0$.
\subsection{Learning Best Responses Against Static Opponents}\label{sec:learning_br}
We first examine whether our method can discover effective strategies against a \textit{static} opponent strategy, which in game-theoretic terms is the problem of finding best responses (\ref{eq:br_defender})--(\ref{eq:br_attacker}). The static strategies are defined in App. \ref{appendix:static_strategies}.
\begin{figure*}
\captionsetup[subfigure]{justification=centering}
  \centering
    \begin{subfigure}[t]{0.6\textwidth}
        \centering
    \scalebox{0.45}{
      \includegraphics{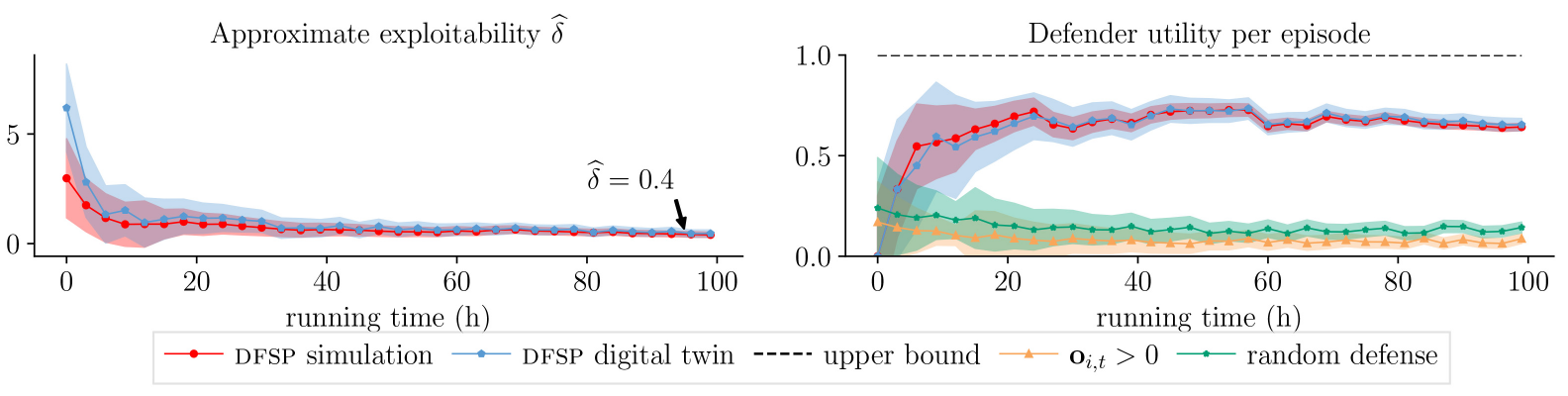}
    }
    \caption{Nash equilibrium learning curves.}
    \label{fig:equilibrium_convergence}
    \end{subfigure}
    \hfill
    \begin{subfigure}[t]{0.37\textwidth}
  \centering
    \scalebox{0.45}{
      \includegraphics{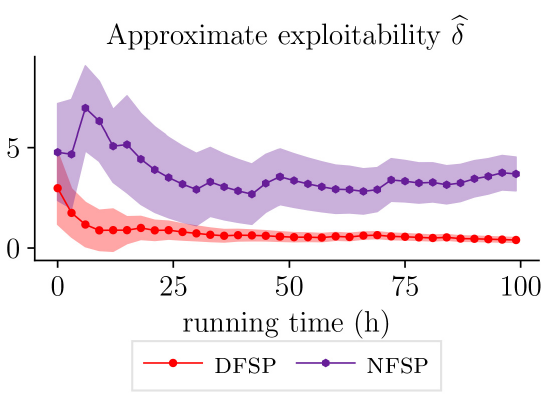}
    }
    \caption{Comparison \dfsp and \nfsp.}\label{fig:nfsp_exp}
    \end{subfigure}
    \caption{Equilibrium learning via \dfsp; the red curves show simulation results and the blue curves show emulation results; the green, orange, purple, and black curves relate to baselines; the figures show approximate exploitability (\ref{eq:approx_exp}) and normalized utility; the curves indicate the mean and the shaded areas indicate the standard deviation over three random seeds.}
    \label{fig:mega_2}
\end{figure*}

To measure the scalability of $\pi^{\text{decomposition}}$ we compare its performance with $\pi^{\text{workflow}}$ and $\pi^{\text{full}}$ on synthetic infrastructures with varying number of nodes $|\mathcal{V}|$ and workflows $|\mathcal{W}|$. To evaluate the optimal stopping approach described in \S \ref{sec:rl_method} we compare its rate of convergence with that of \ppo. Figure \ref{fig:best_response_ppo_returns} shows the learning curves. The red, purple, and pink curves represent the results obtained with $\pi^{\text{decomposition}}$; the blue and beige curves represent the results obtained with $\pi^{\text{workflow}}$; the orange and green curves represent the results obtained with $\pi^{\text{full}}$; and the dashed black lines relate to the baseline strategy that assumes prior knowledge of the opponent's strategy.

We note in Fig. Fig. \ref{fig:best_response_ppo_returns} that all the learning curves of $\pi^{\text{decomposition}}$ converge near the dashed black lines, which suggests that the learned strategies are close to best responses. In contrast, the learning curves of $\pi^{\text{workflow}}$ and $\pi^{\text{full}}$ do not converge near the dashed black lines within the measured time. This is expected as $\pi^{\text{workflow}}$ and $\pi^{\text{full}}$ can not be parallelized like $\pi^{\text{decomposition}}$. (The speedup of parallelization is shown in Fig. \ref{fig:decomposition_scale}.) Lastly, we note in the rightmost plot of Fig. \ref{fig:best_response_ppo_returns} that the optimal stopping approach, which exploits the statement in Thm. \ref{thm:decomp}.C, converges significantly faster than \ppo. An example of a learned optimal stopping strategy based on the linear approximation in (\ref{eq:linear_approx}) is shown in Fig. \ref{fig:learned_switches}.
\subsection{Learning Equilibrium Strategies through Self-Play}
Figures \ref{fig:equilibrium_convergence}--\ref{fig:nfsp_exp} show the learning curves of the strategies obtained during the \dfsp self-play process and the baselines introduced above. The red curves represent the results from the simulator; the blue curves show the results from the digital twin; the green curve give the performance of the random baseline; the orange curve relate to the $\mathbf{o}_{i,t}>0$ baseline; and the dashed black line gives the performance of the baseline strategy that assumes prior knowledge of the attacker actions.

We note that all learning curves in Fig. \ref{fig:equilibrium_convergence} converge, which suggests that the learned strategies converge as well. Specifically, we observe that the approximate exploitability (\ref{eq:approx_exp}) of the learned strategies converges to small values (left plot), which indicates that the learned strategies approximate a Nash equilibrium both in the simulator and in the digital twin. Further, we see from the middle plot that both baseline strategies show decreasing performance as the attacker updates its strategy. In contrast, the defender strategy learned through \dfsp improves its performance over time. This shows the benefit of a game-theoretic approach where the defender strategy is optimized against a dynamic attacker.

Figure \ref{fig:nfsp_exp} compares \dfsp with \nfsp on the simulator. \nfsp implements fictitious self-play and can thus be compared with \dfsp with respect to approximate exploitability (\ref{eq:approx_exp}). We observe that \dfsp converges significantly faster than \nfsp. The fast convergence of \dfsp in comparison with \nfsp is expected as \dfsp is parallelizable while \nfsp is not.
\subsection{Discussion of the Evaluation Results}\label{sec:discussion}
In this work, we propose a formal framework based on recursive decomposition for solving the intrusion response use case, which we evaluate experimentally on a digital twin. The key findings can be summarized as follows.
\begin{enumerate}[(i)]
\item Our framework approximates optimal defender strategies for a practical IT infrastructure (see Fig. \ref{fig:equilibrium_convergence}). While we have not evaluated the learned strategies on the target infrastructure due to safety reasons, the fact that they achieve almost the same performance on the digital twin as on the simulator gives us confidence in the strategies' expected performance on the target infrastructure.

\item Decomposition provides a scalable approach to automate intrusion response for IT infrastructures (see Fig. \ref{fig:best_response_ppo_returns} and Fig. \ref{fig:nfsp_exp}). The intuition behind this finding is that decomposition allows to design efficient divide-and-conquer algorithms that can be parallelized (see Thm. \ref{thm:decomp}.A--B and Alg. \ref{alg:decomposition}).

\item The theory of optimal stopping provides insight about optimal defender strategies, which enables efficient computation of best responses (see the rightmost plot in Fig. \ref{fig:best_response_ppo_returns}). This property can be explained by the threshold structures of the best response strategies, which drastically reduce the search space of possible strategies (Thm. \ref{thm:decomp}.C).

\item Static defender strategies' performance deteriorate against a dynamic attacker whereas defender strategies learned through \dfsp improve over time (see the right plot in Fig. \ref{fig:equilibrium_convergence}). This finding is consistent with previous studies that use game-theoretic approaches (e.g., \cite{quanyan_decompose} and \cite{hammar_stadler_game_23}) and suggests fundamental limitations of static intrusion response systems, such as the Snort \idps.
\end{enumerate}
\section{Conclusions and Future Work}\label{sec:conclusions}
We include elements of game theory and reinforcement learning in a framework to address the problem of automated intrusion response for a realistic use case. We formalize the use case as a partially observed stochastic game. We prove a decomposition theorem stating that the game decomposes recursively into subgames that can be solved efficiently and in parallel, and that the best response defender strategies exhibit threshold structures. This decomposition provides us with a scalable approach to learn near-optimal defender strategies. We develop Decompositional Fictitious Self-Play (\dfsp) -- a fictitious self-play algorithm for finding Nash equilibria. To assess the learned strategies for a target infrastructure, we evaluate them on a digital twin. The results demonstrate that \dfsp converges in reasonable time to near-optimal strategies, both in simulation and on the digital twin, while a state-of-the-art algorithm makes little progress toward an optimal strategy within the same time frame.
\section{Acknowledgments}
The authors would like to thank Pontus Johnson and Quanyan Zhu for useful inputs to this research. The authors are also grateful to Forough Shahab Samani and Xiaoxuan Wang for their constructive comments on a draft of this paper.
\appendices
\section{Hyperparameters and Game Instantiation}\label{appendix:hyperparameters}
We instantiate $\Gamma$ (\ref{def:game}) for the experimental evaluation as follows. Client arrivals are sampled from a stationary Poisson process $Po(\lambda=50)$ and service times are exponentially distributed with mean $\mu=4$. In addition to migrate a node, the defender can shut it down or redirect its traffic to a honeynet, which we model with the zones $\mathfrak{S},\mathfrak{R} \in \mathcal{Z}$. A node $i \in \mathcal{V}$ is shutdown if $v^{(\mathrm{Z})}_{i,t}=\mathfrak{S}$ and have its traffic redirected if $v^{(\mathrm{Z})}_{i,t}=\mathfrak{R}$. The set of local attacker actions is $\mathcal{A}_{\mathrm{A}}^{(\mathrm{V})}=\{\bot, \text{reconnaissance}, \text{brute-force}, \text{exploit}\}$, which we encode as $\{0,1,2,3\}$. These actions have the following effects on the state $\mathbf{s}_{t}$: $\mathbf{a}_{i,t}^{(\mathrm{A})}=1 \implies v^{(\mathrm{R})}_{i,t}=1$, $\mathbf{a}_{i,t}^{(\mathrm{A})}=2 \implies v^{(\mathrm{I})}_{i,t}=1$ with probability $0.3$, and $\mathbf{a}_{i,t}^{(\mathrm{A})}=3 \implies v^{(\mathrm{I})}_{i,t}$ with probability $0.4$. We enforce a tree structure on the target infrastructure in Fig. \ref{fig:use_case} by disregarding the redundant edges in the \rnd zone. The remaining parameters are listed in Table \ref{tab:hyperparams}.

\begin{table}
\centering
\resizebox{1\columnwidth}{!}{%
  \begin{tabular}{ll} \toprule
  \textbf{Game parameters} & {\textbf{Values}} \\
    \hline
    $u_{\mathbf{w},t}$, $\mathcal{A}^{(\mathrm{V})}_{\mathrm{D}}$  & $\sum_{i \in \mathcal{V}_{\mathbf{w}}}[\mathrm{gw} \rightarrow_{t}i]$, $\mathcal{Z} \cup \{\text{access control},\bot\}$ \\
    $|\mathcal{O}^{(\mathrm{V})}|$, $\gamma$, $\eta$, $|\mathcal{Z}|$, $|\mathcal{W}|$, $|\mathcal{V}|$ & $10^3$, $0.9$, $0.4$, $6$, $10$, $64$ \\
     $u^{(\mathrm{w})}_{i}(\bot, l)$, $u^{(\mathrm{w})}_{i}(\mathfrak{S}, l)$,$u^{(\mathrm{w})}_{i}(\mathfrak{R}, l)$, $u^{(\mathrm{w})}_{i}(2, l)$& $0$, $10 + l$, $15+l$, $0.1+l$\\
    $u^{(\mathrm{w})}_{i}(3, l)$, $u^{(\mathrm{w})}_{i}(4, l)$, $u^{(\mathrm{w})}_{i}(5, l)$, $u^{(\mathrm{w})}_{i}(0.8, l)$ & $0.5+l$, $1+l$, $1.5+l$, $2+l$\\
    topology $\mathcal{G}$ and $\mathbf{s}_1^{(\mathrm{D})}$ & see Fig. \ref{fig:use_case} \\
    $|\mathcal{V}_{\mathbf{w}_1}|$,$|\mathcal{V}_{\mathbf{w}_2}|$,$|\mathcal{V}_{\mathbf{w}_3}|$,$|\mathcal{V}_{\mathbf{w}_4}|$,$|\mathcal{V}_{\mathbf{w}_5}|$,$|\mathcal{V}_{\mathbf{w}_6}|$ & $16$, $16$, $16$, $16$, $6$, $4$  \\
    $|\mathcal{V}_{\mathbf{w}_7}|$,$|\mathcal{V}_{\mathbf{w}_8}|$,$|\mathcal{V}_{\mathbf{w}_9}|$,$|\mathcal{V}_{\mathbf{w}_{10}}|$ & $6$, $4$, $6$, $6$  \\
  {\textbf{\textsc{PPO} parameters}} &  \\
  \hline
  lr $\alpha$, batch, \# layers, \# neurons, clip $\epsilon$ & $10^{-5}$, $4\cdot 10^{3}t$, $4$, $64$, $0.2$\\
  GAE $\lambda$, ent-coef, activation & $0.95$, $10^{-4}$, ReLU \\
  {\textbf{NFSP parameters}} &  \\
  \hline
  lr RL, lr SL, batch, \# layers,\# neurons, $\mathcal{M}_{RL}$ & $10^{-2}$, $5\cdot 10^{-3}$, $64$, $2$,$128$, $2\times 10^{5}$ \\
  $\mathcal{M}_{SL}$,$\epsilon$, $\epsilon$-decay, $\eta$ & $2\times 10^{6}$, $0.06$, $0.001$, $0.1$\\
  {\textbf{Stochastic approximation parameters}} &  \\
  \hline
    $c, \epsilon, \lambda, A, a, N, \delta$ & $10$, $0.101$, $0.602$, $100$, $1$, $50$, $0.2$\\
  \bottomrule\\
\end{tabular}
}
\caption{Hyperparameters ($[\cdot]$ is the Iverson bracket).}\label{tab:hyperparams}
\end{table}
\section{Static Defender and Attacker Strategies}\label{appendix:static_strategies}
The static defender and attacker strategies for the evaluation described in \S \ref{sec:learning_br} are defined in (\ref{eq:static_defender})--(\ref{eq:static_attacker}). (w.p is short for "with probability''.)
\begin{align}
&\pi_{\mathrm{D}}(\mathbf{h}_{t}^{(\mathrm{D})})_{i} =
  \begin{dcases}
    \bot & \text{w.p }0.95\\
    j \in \mathcal{Z} & \text{w.p }\frac{0.05}{|\mathcal{Z}|+1}
  \end{dcases}\label{eq:static_defender}\\
&\pi_{\mathrm{A}}(\mathbf{h}_{t}^{(\mathrm{A})})_{i} =
  \begin{dcases}
    \bot & \text{ if }v^{(\mathrm{I})}_{i,t}=1\\
    \bot & \text{w.p }0.8 \text{ if }v^{(\mathrm{R})}_{i,t}=0\\
    \bot & \text{w.p }0.7 \text{ if }v^{(\mathrm{R})}_{i,t}=1,v^{(\mathrm{I})}_{i,t}=0\\
    \text{recon} & \text{w.p }0.2 \text{ if }v^{(\mathrm{R})}_{i,t}=0\\
    \text{brute} & \text{w.p }0.15 \text{ if }v^{(\mathrm{R})}_{i,t}=1,v^{(\mathrm{I})}_{i,t}=0\\
    \text{exploit} & \text{w.p }0.15 \text{ if }v^{(\mathrm{R})}_{i,t}=1,v^{(\mathrm{I})}_{i,t}=0
\end{dcases}\label{eq:static_attacker}
\end{align}
\section{Configuration of the Infrastructure in Fig. \ref{fig:use_case}}\label{appendix:infrastructure_configuration}
The configuration of the target infrastructure (Fig. \ref{fig:use_case}) is available in Tables \ref{tab:target_infra_config} and \ref{tab:workflows}.

\begin{table*}
\centering
\resizebox{1\linewidth}{!}{%
\begin{tabular}{llllll} \toprule
  {\textit{ID(s)}} & {\textit{Type}} & {\textit{Operating system}} & {\textit{Zone}} & {\textit{Services}} & {\textit{Vulnerabilities}} \\ \midrule
  $1$ & Gateway & \ubuntu 20 & - & \snort (ruleset v2.9.17.1), \ssh, \openflow v1.3, \ryu \sdn controller & -\\
  $2$ & Gateway & \ubuntu 20 & \dmz & \snort (ruleset v2.9.17.1), \ssh, \ovs v2.16, \openflow v1.3 & -\\
  $28$ & Gateway & \ubuntu 20 & \rnd & \snort (ruleset v2.9.17.1), \ssh, \ovs v2.16, \openflow v1.3 & -\\
  $3$,$12$ & Switch & \ubuntu 22 & \dmz & \ssh, \openflow v1.3 , \ovs v2.16& -\\
  $21,22$ & Switch & \ubuntu 22 & - & \ssh, \openflow v1.3, \ovs v2.16 & -\\
  $23$ & Switch & \ubuntu 22 & \admin & \ssh, \openflow v1.3, \ovs v2.16 & -\\
  $29$-$48$ & Switch & \ubuntu 22 & \rnd & \ssh, \openflow v1.3, \ovs v2.16 & -\\
  $13$-$16$ & Honeypot & \ubuntu 20 & \dmz & \ssh, \snmp, \postgres, \ntp & -\\
  $17$-$20$ & Honeypot & \ubuntu 20 & \dmz & \ssh, \irc, \snmp, \ssh, \postgres & -\\
  $4$ & App node & \ubuntu 20 & \dmz & \http, \dns, \ssh & \cwe-1391\\
  $5$, $6$ & App node & \ubuntu 20 & \dmz & \ssh, \snmp, \postgres, \ntp & -\\
  $7$ & App node & \ubuntu 20 & \dmz & \http, \telnet, \ssh & \cwe-1391\\
  $8$ & App node & \debian \jessie & \dmz & \ftp, \ssh, \apache 2,\snmp & \cve-2015-3306\\
  $9$,$10$ & App node & \ubuntu 20 & \dmz & \ntp, \irc, \snmp, \ssh, \postgres & -\\
  $11$ & App node & \debian \jessie & \dmz & \apache 2, \smtp, \ssh & \cve-2016-10033\\
  $24$ & Admin system & \ubuntu 20 & \admin & \http, \dns, \ssh & \cwe-1391\\
  $25$ & Admin system & \ubuntu 20 & \admin & \ftp, \mongo, \smtp, \tomcat, \ts 3, \ssh & -\\
  $26$ & Admin system & \ubuntu 20 & \admin & \ssh, \snmp, \postgres, \ntp & -\\
  $27$ & Admin system & \ubuntu 20 & \admin & \ftp, \mongo, \smtp, \tomcat, \ts 3, \ssh & \cwe-1391\\
  $49$-$59$ & Compute node & \ubuntu 20 & \rnd & \spark, \hdfs & -\\
  $60$ & Compute node & \debian \wheezy & \rnd & \spark, \hdfs, \apache 2,\snmp, \ssh & \cve-2014-6271\\
  $61$ & Compute node & \debian 9.2 & \rnd & \irc, \apache 2, \ssh & \cwe-89\\
  $62$ & Compute node & \debian \jessie & \rnd & \spark, \hdfs, \ts 3, \tomcat, \ssh & \cve-2010-0426\\
  $63$ & Compute node & \debian \jessie & \rnd & \ssh, \spark, \hdfs & \cve-2015-5602\\
  $64$ & Compute node & \debian \jessie & \rnd & \samba, \ntp, \ssh, \spark, \hdfs & \cve-2017-7494\\
  \bottomrule\\
\end{tabular}
}
\caption{Configuration of the target infrastructure shown in Fig. \ref{fig:use_case}; each row contains the configuration of one or more nodes; vulnerabilities in specific software products are identified by the vulnerability identifiers in the Common Vulnerabilities and Exposures (\cve) database \cite{cve}; vulnerabilities that are not described in the \cve database are categorized according to the types of the vulnerabilities they exploit based on the Common Weakness Enumeration (\cwe) list \cite{cwe}.}\label{tab:target_infra_config}
\end{table*}

\begin{table}
\centering
\resizebox{1\columnwidth}{!}{%
\begin{tabular}{llll} \toprule
  {\textit{ID}} & {\textit{Name}} & {\textit{Zone}} & {\textit{Nodes}} \\ \midrule
  1 & \spark $1$ & \rnd & $1$, $21$, $22$, $28$, $(29-32)$, $(33-34)$, $(41-42)$, $(49-52)$ \\
  2 & \spark $2$ & \rnd & $1$, $21$, $22$, $28$, $(29-32)$, $(35-36)$, $(43-44)$, $(53-56)$ \\
  3 & \spark $3$ & \rnd & $1$, $21$, $22$, $28$, $(29-32)$, $(37-38)$, $(45-46)$, $(57-60)$ \\
  4 & \spark $4$ & \rnd & $1$, $21$, $22$, $28$, $(29-32)$, $(39-40)$, $(47-48)$, $(61-65)$ \\
  5 & Web $1$ & \dmz & $1$, $2$, $3$, $4$, $5$, $6$ \\
  6 & Web $2$ & \dmz & $1$, $2$, $3$, $7$ \\
  7 & Storage $1$ & \dmz & $1$, $2$, $3$, $8$, $9$, $10$ \\
  8 & Mail $1$ & \dmz & $1$, $2$, $3$, $11$ \\
  9 & Admin $1$ & \admin & $1$, $21$, $22$, $23$, $24$, $25$ \\
  10 & Admin $2$ & \admin & $1$, $21$, $22$, $23$, $25$, $26$ \\
  \bottomrule\\
\end{tabular}
}
\caption{Workflows of the target infrastructure (Fig. \ref{fig:use_case}).}\label{tab:workflows}
\end{table}

\section{Attacker Actions}\label{appendix:attacker actions}
The attacker actions and their descriptions are listed in Table \ref{tab:attacker_actions_descr}.
\begin{table}
\centering
\resizebox{1\columnwidth}{!}{%
\begin{tabular}{ll} \toprule
  {\textit{Action}} & {\textit{Description}} \\ \midrule
  \tcpp scan & \tcpp port scan by sending \syn or empty packets using \texttt{nmap}, \\
             & it allows detecting open \tcpp ports by checking the responses\\
  \udp port scan & \udp port scan by sending \udp packets using \texttt{nmap}, \\
             & it allows detecting open \udp ports by checking the responses\\
  ping scan & IP scan with \icmp ping messages, it allows detecting alive hosts \\
  \vulscan & vulnerability scan using \texttt{nmap}, it allows detecting known vulnerabilities\\
  brute-force attack & performs a dictionary attack against a login service using \texttt{nmap}\\
  \cve-2017-7494 exploit & uploads a malicious binary to the \samba service for remote code execution\\
  \cve-2015-3306 exploit & uses the \texttt{mod\_copy} in \texttt{proftpd} for remote code execution  \\
  \cve-2014-6271 exploit & uses a vulnerability in \texttt{bash} for remote code execution \\
  \cve-2016-10033 exploit & uses \texttt{phpmailer} for remote code execution \\
  \cve-2015-1427 exploit & uses the scripting engine in \texttt{elasticsearch} for remote code execution \\
  exploit of \cwe-89 weakness on DVWA \cite{dvwa} & sends a HTTP request injecting SQL code for SQLite3 \\
  \bottomrule\\
\end{tabular}
}
\caption{{Descriptions of the attacker actions; shell commands and scripts for executing the actions are listed in \cite{supp_material_hammar_stadler}}.}\label{tab:attacker_actions_descr}
\end{table}

\bibliographystyle{IEEEtran}
\bibliography{references,url}
\end{document}